\documentclass[useAMS,usenatbib]{mn2e}
\usepackage{amsmath,graphicx,txfonts}
\defcitealias{har02}{Paper I}
\defcitealias{har06}{Paper II}
\defcitealias{mcl07}{Paper IV}

\title[Two Globular Clusters in the M81 Group]
{Ages and structural and dynamical parameters of two globular clusters in the M81 group}
\author[J. Ma et al.]
{Jun Ma,$^{1,2}$\thanks{E-mail: majun@nao.cas.cn} Song Wang,$^{1}$\thanks{E-mail: somgw@bao.ac.cn} Zhenyu Wu,$^{1,2}$ TianMeng Zhang,$^{1,2}$ Hu Zou,$^{1}$ Zhimin Zhou,$^1$
\newauthor
Jundan Nie,$^1$ Xu Zhou,$^1$ Xiyang Peng,$^1$ Jiali Wang,$^1$ Jianghua Wu,$^3$ Cuihua Du$^4$
\newauthor
and Qirong Yuan$^5$\\
$^1$Key Laboratory of Optical Astronomy, National Astronomical Observatories, Chinese Academy of Sciences, Beijing 100012, China\\
$^2$College of Astronomy and Space Sciences, University of Chinese Academy of Sciences, Beijing 100049, China\\
$^3$Department of Astronomy, Beijing Normal University, Beijing 100875, China\\
$^4$School of Physics Sciences, University of the Chinese Academy of Sciences, Beijing 100049, China\\
$^5$Department of Physics, Nanjing Normal University, WenYuan Road 1, Nanjing 210046, China}
\date{Received; Accepted}
\pubyear{468, 4513--4528, 2017}
\begin{document}
\label{firstpage}
\maketitle

\begin{abstract}

GC-1 and GC-2 are two globular clusters (GCs) in the remote halo of M81 and M82 in the M81 group
discovered by Jang et al. using the {\it Hubble Space Telescope} ({\it HST}) images. These two GCs
were observed as part of the Beijing--Arizona--Taiwan--Connecticut (BATC) Multicolor Sky Survey,
using 14 intermediate-band filters covering a wavelength range of 4000--10000 \AA.
We accurately determine these two clusters' ages and masses by comparing their spectral energy distributions (from 2267 to
20000~{\AA}, comprising photometric data in the near-ultraviolet of the {\it Galaxy Evolution Explorer},
14 BATC intermediate-band, and Two Micron All Sky Survey near-infrared $JHK_{\rm s}$ filters) with
theoretical stellar population-synthesis models, resulting in ages of $15.50\pm3.20$ for GC-1 and
$15.10\pm2.70$ Gyr for GC-2. The masses of GC-1 and GC-2 obtained here are $1.77-2.04\times 10^6$
and $5.20-7.11\times 10^6 \rm~M_\odot$, respectively. In addition, the deep observations with the Advanced
Camera for Surveys and Wide Field Camera 3 on the {\it HST} are used to provide the surface brightness
profiles of GC-1 and GC-2. The structural and dynamical parameters are derived from fitting the profiles
to three different models; in particular, the internal velocity dispersions of GC-1 and GC-2 are derived,
which can be compared with ones obtained based on spectral observations in the future. For the first time,
in this paper, the $r_h$ versus $M_V$ diagram shows that GC-2 is an ultra-compact dwarf in the M81 group.

\end{abstract}


\section{Introduction}
\label{sec:intro}

The study of globular clusters (GCs) plays an important role in our understanding of
the evolution and history of galaxies. The Galactic GCs, the stars of which are thought
to be among the oldest objects in the universe, can provide important information regarding
the minimum age of the universe and the early formation history of our Galaxy. In addition,
studying the spatial structures and dynamics of GCs is of great importance for understanding
both their formation conditions and dynamical evolution within the tidal fields of their
galaxies \citep{barmby07}.

The most direct method to determine a cluster's age is by employing main-sequence photometry,
since the absolute magnitude of the main-sequence turn-off is predominantly affected by age
\citep[see][and references therein]{puzia02b}. However, this method was mainly applied to
the star clusters in the Milky Way and its satellites \citep[e.g.][]{rich01}. Generally,
the ages of extragalactic star clusters are determined by comparing their observed spectral
energy distributions (SEDs) and/or spectroscopy with the predictions of simple stellar
population (SSP) models \citep[see][and references therein]{Ma10}.

The structural and dynamical parameters of clusters are often determined by fitting the
surface brightness profiles to a number of different models, combined with mass-to-light
ratios ($M/L$ values) estimated from velocity dispersions or population-synthesis models.
We only mention three models that will be used in this paper. The first is based on a
single-mass, isotropic, modified isothermal sphere developed by \citet{king66}. The second is
a further modification of a single-mass, isotropic isothermal sphere based on the ad hoc
stellar distribution function of \citet{wilson75}. The third model is based on the $R^{1/n}$
surface density profile of \citet{sersic68}. Using the three models, some authors have achieved
some success in determining structural and dynamical parameters of clusters in the Local galaxies:
the Milky Way, the Large and Small Magellanic Clouds, Fornax and Sagittarius dwarf spheroidal
galaxies \citep{mm05}, M31 \citep{barmby07,barmby09,wang13}, NGC 5128 \citep{mclau08}, and M33
\citep{ma15}.

Except for the Local Group galaxies, the M81 group includes the nearest large spirals outside
the Local Group. M81 locates in the centre of the M81 group, and 36 member galaxies surround M81
\citep{Chiboucas13}. Although the GC system of M81 has come under recent detailed scrutiny
\citep[see][and references therein]{Ma13}, however structural and dynamical studies of the GC
system of the M81 group are very limited: \citet{cft01} estimated the core radii of 114 compact
star clusters of M81 based on the {\it Hubble Space Telescope} ({\it HST})/Wide Field Planetary
Camera 2 imaging; \citet{N11} estimated the effective radii of 419 GC candidates of M81 using
{\it HST}/Advanced Camera for Surveys (ACS) imaging; and \citet{lim13} estimated the effective
radii of 1105 star clusters in M82 using {\it HST}/ACS imaging. \citet{Jang12} and \citet{Mayya13}
determined structural parameters for three GCs in the M81 group by fitting the King models to the
surface brightness profiles of the {\it HST} images. However, for two of them (i.e. GC-1 and GC-2),
\citet{Jang12} only derived their core and half-light radii. GC-1 and GC-2 are of interest because
they are in the remote haloes of M81 and M82 in the M81 group recently discovered by \citet{Jang12};
in particular, GC-2 is the most isolated GC in the local universe. So, we will derive the structural
and dynamical parameters of GC-1 and GC-2. In addition, we will also accurately determine these two
clusters' ages and masses by comparing their SEDs with theoretical stellar population-synthesis models.

The distances were taken to be $(m-M)_0=27.80\pm0.03$ to GC-1 and $(m-M)_0=28.04\pm0.04$ to GC-2
throughout, which were derived by \citet{Jang12}.

In this paper, we will determine the ages and masses, and structural and dynamical parameters, for
GC-1 and GC-2. We will describe the details of the observations and our approach to the data reduction
with the Beijing--Arizona--Taiwan--Connecticut (BATC) system in Section 2 and with the {\it HST} programs
in Sections 2 and 4. We will determine the ages and masses of GC-1 and GC-2 by comparing observational
SEDs with population-synthesis models in Section 3, and determine structural and dynamical parameters of
GC-1 and GC-2 in Section 5. We make comparison to previous results and discussion in Section 6, and provide a
summary in Section 7.

\section{Archival images of BATC, 2MASS, {\it GALEX} and {\it HST}}
\label{sec:data}

In this section, we will determine the magnitudes of GC-1 and GC-2 based on the archival images of
the BATC Multicolor Sky Survey, Two Micron All Sky Survey (2MASS) and {\it Galaxy Evolution Explorer}
({\it GALEX}) using a standard aperture photometry approach, i.e. the {\sc phot} routine in
{\sc daophot} \citep{stet87}. In addition, we will also describe {\it HST} observations as well as
magnitude transformation in this section.

\subsection{Intermediate-band photometry}

The M81 field is part of a galaxy calibration program of the BATC Multicolor Sky Survey. The BATC
program uses the 60/90 cm Schmidt Telescope at the Xinglong Station of the National Astronomical
Observatories, Chinese Academy of Sciences. This system includes 15 intermediate-band filters,
covering a range of wavelength from 3000 to 10000 \AA~\citep[see][for details]{fan96}. In order to
study the stellar populations of GC-1 and GC-2, we extracted 322 images of M81 field as part of the
BATC Multicolor Sky Survey, taken in 14 intermediate-band filters with a total exposure time of
$\sim 100$ h from February 5, 1995 to April 30, 2002. These images were observed in the good nights
including the photometric nights. The images of one filter were observed as possible as
in the same night. The dome flat-field images were taken by using an isotropic diffuser right
in front of the Schmidt corrector plate (the entrance pupil of the telescope) \citep[see][for details]{zhou04}. The images were reduced
with standard procedures, including bias subtraction and flat-fielding of the CCD images, with an automatic data reduction software named
{\sc pipeline i}, developed for the BATC Multicolor Sky Survey (Fan et al. 1996; Zheng et al. 1999). Also in {\sc pipeline i}, the astrometric plate
solution is obtained by registering the brighter stars in each image with the {\it HST} Guide Star Catalogue (GSC)
coordinate system (Jenkner et al. 1990). The positions of the stars in the images ($x$ and $y$) then can be easily transformed to
the equatorial coordinates ($\alpha$ and $\beta$) in J2000. The final RMS errors in positional accuracy of the stars are about 0.5 arcsec. Then, the multiple images
of the same filter were combined to improve the signal-to-noise ratios (SNRs), and the cosmic rays and bad pixels were corrected
by comparison of multiple images during combination. Before the combination, the other images were shifted and rotated according to one image.
The images observed in different exposures were then stacked in the combination.

\begin{figure*}
\centerline{\hfil
   \includegraphics[width=85mm]{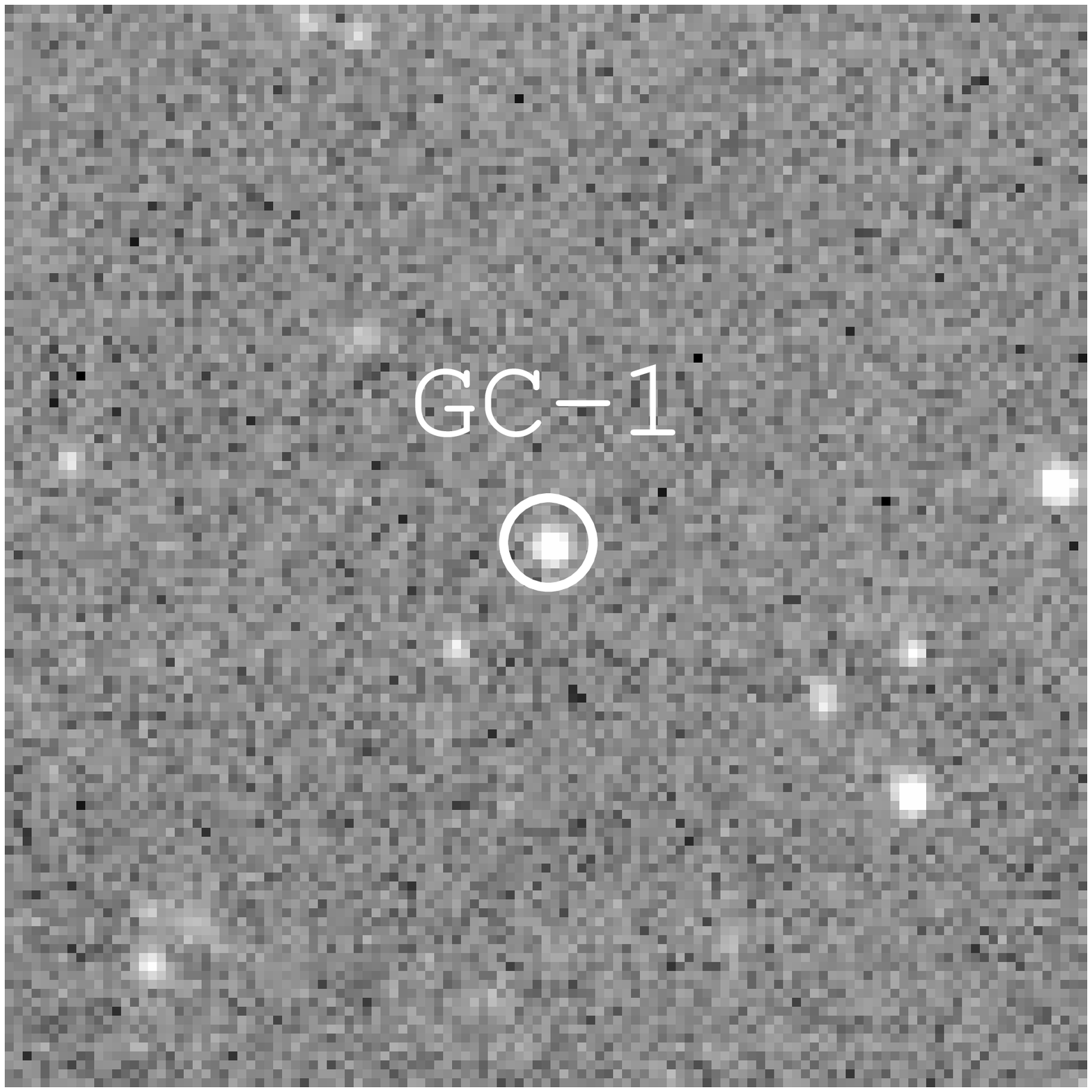}
   \hfill
   \includegraphics[width=85mm]{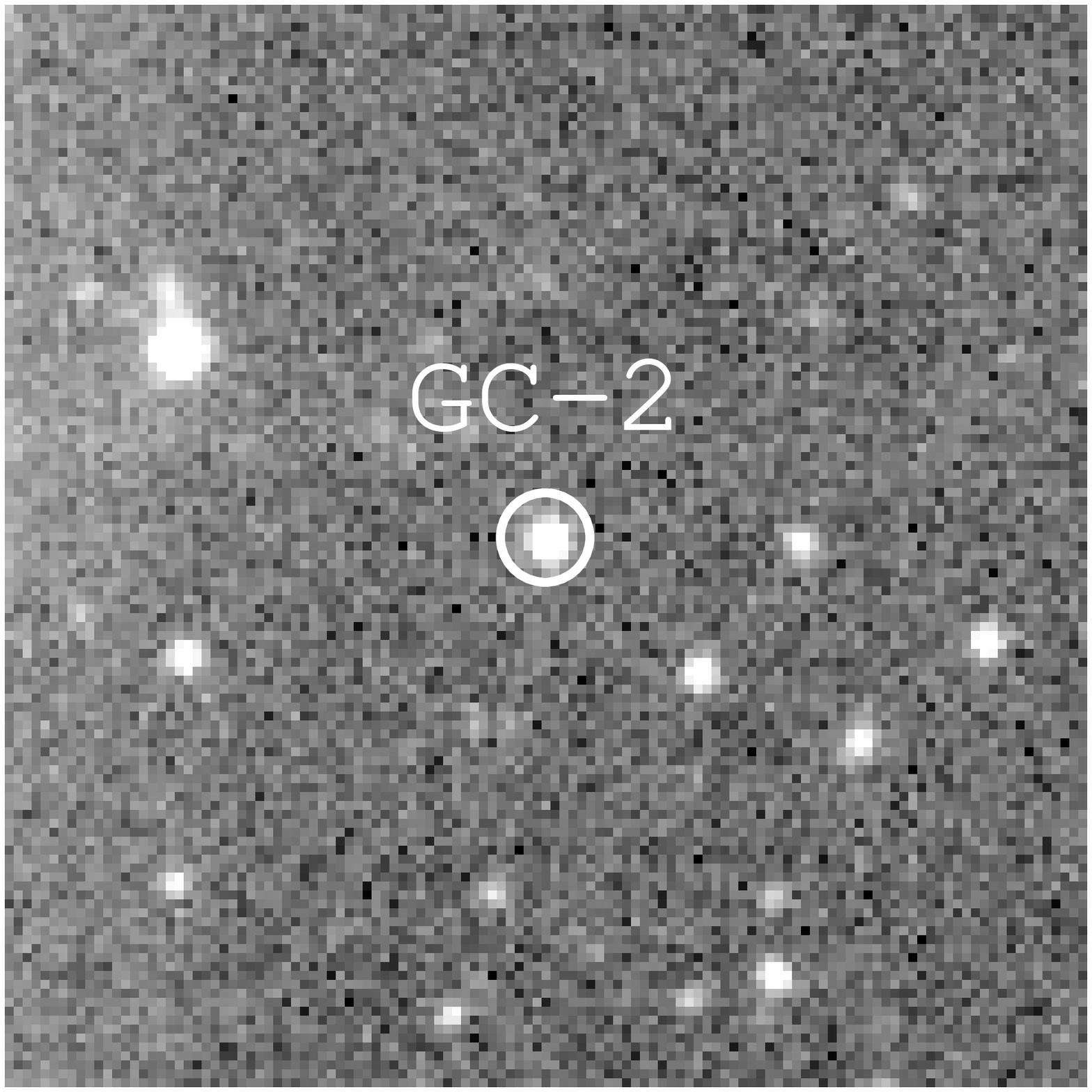}
\hfil}
\caption{Images of GC-1 and GC-2 in the BATC $g$ band obtained with the 60/90 cm Schmidt telescope. The circles are photometric apertures used in this paper.}
\end{figure*}

Calibration of the magnitude zero level in the BATC photometric system is similar to that of the
spectrophotometric AB magnitude system. For flux calibration, the Oke-Gunn \citep{ok83} primary
flux standard stars HD 19445, HD 84937, BD +26$^{\circ}$2606 and BD +17$^{\circ}$4708 were observed
during photometric nights \citep{yan00}. On the photometric nights, two or more standard stars
as well as the BATC programmed fields were observed between air masses of 1 and 2. The standard stars are
observed as frequently as possible using the central part of the CCD (size $300\times300$) for saving
readout time and disc space. The extinction coefficient and magnitude zero obtained from the
standard stars are used for making the flux calibration on the BATC programmed field images.
The exposure times of the BATC programmed fields are 300 s (i.e. in short exposures).
Using the images of the standard stars observed on
the photometric nights, the extinction coefficients at any given time in a night
and the zero point of the instrumental magnitude are obtained. The instrumental magnitudes $M_{\rm inst}$
of selected bright, isolated stars in the BATC programmed field images observed in short exposures
can be readily transformed to the BATC AB magnitude system $M_{\rm batc}$,
and then these bright, isolated stars were taken as secondary standards and were subsequently used
to perform calibration on the combined image.
Table 1 lists the parameters of the BATC Multicolor Filter System and the statistics of observations.
Column 7 of Table 1 gives the errors of flux calibration in each filter.

Fig. 1 shows the finding charts of GC-1 and GC-2 in the BATC $g$ band.

\begin{table}
\caption{Parameters of the BATC filters and statistics of observations for M81 field.}\label{table1}
\begin{tabular}{ccccccc}
\hline
No. & Name & cw (\AA)${^a}$ & FWHM (\AA)${^b}$ & Exp.
(h)${^c}$ & N.img${^d}$ & rms${^f}$\\
\hline
1  & $b$ & 3890   & 291 & 03:40 & 11 & 0.009\\
2  & $c$ & 4210   & 309 & 03:53 & 14 & 0.004\\
3  & $d$ & 4546   & 332 & 12:20 & 39 & 0.013\\
4  & $e$ & 4872   & 374 & 06:10 & 21 & 0.005\\
5  & $f$ & 5250   & 344 & 06:05 & 19 & 0.005\\
6  & $g$ & 5785   & 298 & 05:12 & 18 & 0.004\\
7  & $h$ & 6075   & 308 & 04:00 & 12 & 0.006\\
8  & $i$ & 6710   & 491 & 06:00 & 18 & 0.006\\
9  & $j$ & 7010   & 238 & 05:20 & 16 & 0.007\\
10 & $k$ & 7530   & 192 & 05:40 & 17 & 0.013\\
11 & $m$ & 8000   & 255 & 05:20 & 16 & 0.006\\
12 & $n$ & 8510   & 167 & 15:00 & 45 & 0.005\\
13 & $o$ & 9170   & 247 & 16:40 & 50 & 0.036\\
14 & $p$ & 9720   & 275 & 08:40 & 26 & 0.039\\
\hline
\end{tabular}

\medskip
$^{a}$ Central wavelength for each BATC filter.

$^{b}$ Passband width for each BATC filter.

$^{c}$ A total exposure time for each BATC filter.

$^{d}$ Image numbers for each BATC filter.

$^{f}$ Calibration error, in magnitude, for each filter as obtained from the standard stars.
\end{table}


To determine the total luminosities of GC-1 and GC-2, we produced curves of growth from the $g$-band
photometry obtained through apertures with radii in the range 2--11 pixel with 1 pixel increments. These
curves of growth were used to determine the aperture size required to enclose the total cluster light.
The most appropriate photometric radius that includes all light from the objects is adopted independently
for each cluster. The local sky background was measured in an annulus with an inner radius which being
larger 1 pixel than photometric radius and $\sim 8.4$ arcsec wide, in which the mode was used. Finally, we
obtained photometry for GC-1 and GC-2 in the individual 14 intermediate-band filters. Table 2 lists our
new magnitudes, with errors given by {\sc iraf/daophot}. Column (1) gives the cluster names. The $1\sigma$
magnitude uncertainties from {\sc daophot} are listed for each object on the second line for the corresponding bands.
Sources of CCD photometric error of the BATC photometric system include: photon statistics; readout noise; random and systematic errors
from bias substraction and flat-fielding, and image defects, and possible non-linearity of the CCD. The analysis
program {\sc daophot} gives the theoretical prediction of the error based on the photon statistics it measures, which
is included in Table 2. \citet{fan96} have analyze the other errors in detail, and found that these errors
are not larger than 0.02 mag for stars brighter than $m=16$ in most filters.

\begin{table*}
\begin{minipage}{170mm}
\scriptsize
\caption{BATC, 2MASS and GALEX photometry of GC-1 and GC-2 in the M81 group.}\label{table2}
\begin{tabular}{cccccccccccccccccccc}
\hline
ID &  $b$ & $c$ & $d$ & $e$  & $f$  & $g$  & $h$ & $i$  & $j$  & $k$  & $m$ & $n$ & $o$ & $p$ & $J$ & $H$ & $K_s$  &  NUV \\
 &  (mag) & (mag) & (mag) & (mag) & (mag) & (mag) & (mag) & (mag) & (mag) & (mag) & (mag) & (mag) & (mag) & (mag) & (mag) & (mag) & (mag) & (mag)\\
\hline
GC-1             &19.715  &19.326  &19.221  &18.989  &18.835  &18.539  &18.454  &18.328  &18.278  &18.106  &17.997  &18.012  &17.915  &17.857  &17.060 &16.449 &...    &22.974\\
                 &0.270   &0.224   &0.039   &0.024   &0.019   &0.035   &0.028   &0.009   &0.030   &0.030   &0.033   &0.038   &0.072   &0.104   &0.029  &0.021  &...    &0.427 \\
GC-2             &18.934  &19.057  &18.260  &18.077  &17.868  &17.617  &17.486  &17.401  &17.315  &17.208  &17.112  &17.108  &16.901  &16.856  &15.897 &15.587 &15.207 &22.134\\
                 &0.106   &0.112   &0.017   &0.011   &0.010   &0.016   &0.011   &0.005   &0.014   &0.015   &0.015   &0.013   &0.028   &0.051   &0.017  &0.014  &0.012  &0.290 \\
\hline
\end{tabular}
\end{minipage}
\end{table*}


\subsection{Near-infrared 2MASS photometry}

We used the 2MASS archival images of GC-1 and GC-2 in the $JHK_{\rm s}$ bands to do photometry. The images
were retrieved using the 2MASS Batch Image Service\footnote{http://irsa.ipac.caltech.edu/applications/2MASS/IM/batch.html.}.
The uncompressed atlas images were used with a resampled spatial resolution of $\sim$$1$ arcsec pixel$^{-1}$.
The photometry routine we used is also {\sc iraf/daophot}
\citep{stet87}.

We determine the total luminosity of each object in the $JHK_{\rm s}$ bands based on the processes in
Section 2.1. We produced a curve of growth from the $J$-band photometry obtained through apertures with
radii in the range of 3--14 pixel with 1 pixel increments. The local sky background was measured in an annulus
with an inner radius 1 pixel larger than the photometric radius and 5 pixels wide. At last, the instrumental
magnitudes were then calibrated using the relevant zero points obtained from the photometric header keywords
of each image. The calibrated photometry of GC-1 and GC-2 in $J$, $H$ and $K_s$ bands is summarized in Table 2,
in conjunction with the $1\sigma$ magnitude uncertainties obtained from {\sc daophot}. The photometry of GC-1 in
$K_s$ cannot be obtained here because of low SNRs. In addition, since the observed magnitudes in the 2MASS
photometric system are given in the Vega system, we transformed them to the AB system for our fits
(see Section 3.3 for details). The photometric offsets in the 2MASS filters between the Vega and AB systems
were obtained on the basis of equations (7) and (8) in the manual provided by \citet[hereafter BC03]{bc03}.

\subsection{{\it GALEX} ultraviolet photometry}

We used the {\it GALEX} archival images of GC-1 and GC-2 in the FUV and NUV bands to do photometry.
The images were retrieved using the {\it GALEX} Batch Image Service\footnote{http://galex.stsci.edu/GR6/.}.
The exposure times were 14706.7 s in the FUV band and 29421.55 s in the NUV band, respectively.
The images were sampled with $1.5''$ pixels. The relevant zero-points for photometry are 20.08 and 18.82 in
the NUV and FUV magnitudes, respectively \citep{Morrissey07}.

We determine the total luminosity of each object in the FUV and NUV bands using the processes in Section 2.1.
We produced a curve of growth from the NUV-band photometry obtained through apertures with radii in the range
of 2--13 pixel with 1 pixel increments. The calibrated photometry of GC-1 and GC-2 in the NUV band is summarized
in Table 2. The photometry of GC-1 and GC-2 in the FUV band cannot be obtained here because of low SNRs. It is
known that the {\it GALEX} photometric system is calibrated to match the spectrophotometric AB system \citep{rey06}
as the BATC photometric system does.

\subsection{{\it HST} observation}

The observations used here come from the {\it HST} program 11613 (P.I.: de Jong), in which GC-1 was observed by the
{\it HST}/ACS/WFC in the $F606W$ and $F814W$ bands, and GC-2 was observed by the {\it HST}/Wide Field Camera 3 (WFC3)/UVIS
in the $F606W$ and $F814W$ bands. GC-1 was observed with a total exposure time of 850 s in the $F606W$ band and 690 s in the
$F814W$ band, and GC-2 was observed with a total exposure time of 735 s in the $F606W$ band and 1225 s in the $F814W$ band.
We obtained the combined drizzled images from the Hubble Legacy Archive. In addition, for easy comparison with catalogues
of the GCs in the Milky Way in the future work, we transform the ACS/WFC and WFC3/UVIS magnitudes in the $F606W$ band to
the standard $V$. \citet{wang12} has presented transformation from ACS/WFC magnitude to standard $V$ magnitude in their
equation (7). However, till now transformations from WFC3/UVIS to standard $V$ magnitude has not been presented in any
references. We used equation (7) of \citet{wang12} for the transformation in this paper, and $V$ and $I$ magnitudes are
obtained in this paper (see Section 3.4 in detail). We rewrite equation (7) of \citet{wang12} below:
\begin{equation}
(V-{F606W})_0=-0.067+0.340(V-I)_0-0.038(V-I)_0^2.
\end{equation}

We will use these $\it HST$ images to derive the structural and dynamical parameters of GC-1 and GC-2 (see Sections 4
and 5 for details).

\section{Stellar population}

\subsection{Metallicities and reddening values}

Cluster SEDs are determined by the combination of their ages and metallicities, which is often referred to as the
age--metallicity degeneracy. Therefore, the age of a cluster can only be constrained accurately if the metallicity
is determined independently with confidence. For GC-1 and GC-2, only \citet{Jang12} determined their metallicities.
These authors presented the color--magnitude diagrams for GC-1 and GC-2, and derived the metallcities to be
$\rm{[Fe/H]}=-2.23\pm0.11$ for GC-1 and $\rm{[Fe/H]}=-2.23\pm0.12$ for GC-2 based on the mean colour of the red
giant branch stars. In addition, \citet{Jang12} derived the matallicity of GC-2 to be $\rm{[Fe/H]}=-2.3\pm0.12$
based on its optical spectrum observed by the Sloan Digital Sky Survey (SDSS). Here we adopted the values of
metallicity from \citet{Jang12}: $\rm{[Fe/H]}=-2.23\pm0.11$ for GC-1 and $\rm{[Fe/H]}=-2.3\pm0.12$ for GC-2.

In order to obtain intrinsic SEDs of GC-1 and GC-2, the photometry must be corrected for reddening. Since GC-1
and GC-2 are in the remote haloes of M81 and M82, only the foreground extinction contribution of the Milky Way is
considered. The dust map of \citet{Schlegel98} presents $E(B-V)=0.09$ for GC-1 and $E(B-V)=0.10$ for GC-2
(see also Jang et al. 2012 and Monachesi et al. 2013).

\subsection{Stellar populations and synthetic photometry}

To determine the ages and masses of GC-1 and GC-2, we compared their SEDs with theoretical stellar population-
synthesis models. The SEDs consist of photometric data in the NUV band of {\it GALEX}, 14 BATC intermediate-band
and 2MASS near-infrared (NIR) $JHK_s$ filters obtained in this paper. GC-1 and GC-2 are both very metal poor
(see Section 3.1). So, we use the SSP models of BC03, which provide the evolution of
the spectra and photometric properties for a wide range of stellar metallicities. For example, BC03 SSP models
based on the Padova-1994 evolutionary tracks include six initial metallicities:
$Z= 0.0001$, 0.0004, 0.004, 0.008, 0.02 $(Z_\odot)$, and 0.05, corresponding to
${\rm [Fe/H]}=-2.25$, $-1.65$, $-0.64$, $-0.33$, $+0.09$, and $+0.56$. In this paper, we adopt the high-resolution
SSP models using the Padova-1994 evolutionary tracks to determine the most appropriate ages for GC-1 and GC-2.
And a \citet{salp55} initial mass function (IMF) is also used. These SSP models contain 221 spectra describing the
spectral evolution of SSPs from $1.0\times10^5$ yr to 20 Gyr. The evolving spectra include the contribution of
the stellar component at wavelengths from 91~\AA~to $160~\mu$m.

Since our observational data are integrated luminosities through a given set of filters, we convolved the
theoretical SSP SEDs of BC03 with the {\it GALEX} NUV, BATC $b-p$ and 2MASS $JHK_{\rm s}$ filter response
curves to obtain synthetic optical and NIR photometry for comparison. The synthetic $i{\rm th}$ filter magnitude
can be computed by
\begin{equation}
m=-2.5\log\frac{\int_{\lambda}F_{\lambda}\varphi_{i} (\lambda){\rm
d}\lambda}{\int_{\lambda}\varphi_{i}(\lambda){\rm d}\lambda}-48.60,
\end{equation}
where $F_{\lambda}$ is the theoretical SED, and $\varphi_{i}$ is the response function of the $i{\rm th}$ filter
of the {\it GALEX}, BATC and 2MASS photometric systems. Here, $F_{\lambda}$ varies with age and metallicity.

\subsection{Fit Results}

We use a $\chi^2$ minimization approach to examine which SSP models are most compatible with the observed SEDs of
GC-1 and GC-2, following

\begin{equation}
\chi^2=\sum_{i=1}^{N}{\frac{[m_{\lambda_i}^{\rm intr}-m_{\lambda_i}^{\rm
mod}(t)]^2}{\sigma_{i}^{2}}},
\end{equation}
where $m_{\lambda_i}^{\rm mod}(t)$ is the integrated magnitude in the $i{\rm th}$ filter of a theoretical SSP at age
$t$, $m_{\lambda_i}^{\rm intr}$ represents the intrinsic integrated magnitude in the same filter, and
\begin{equation}
\sigma_i^{2}=\sigma_{{\rm obs},i}^{2}+\sigma_{{\rm mod},i}^{2}+\sigma_{{\rm md},i}^{2}.
\end{equation}
Here, $\sigma_{{\rm obs},i}$ is the observational uncertainty from Table 2 of this paper, $\sigma_{{\rm mod},i}$ is
the uncertainty associated with the model itself, and $\sigma_{{\rm md},i}$ is associated with the uncertainty with
the distance modulus adopted here. Following \citet{ma12}, we adopt $\sigma_{{\rm mod},i}=0.05$ mag.
For $\sigma_{{\rm md},i}$, we adopt 0.03 and 0.04 for GC-1 and GC-2, respectively, which are from \citet{Jang12}.
$N$ is the number of photometries used in the fitting.

Before fitting, we obtained the theoretical SEDs for the metallicity $\rm [Fe/H]=-2.23$ model by interpolation of
between ${\rm [Fe/H]}=-2.25$ and $-1.65$ models for GC-1. For GC-2, we used the theoretical SEDs for the metallicity
$\rm [Fe/H]=-2.25$ model for fitting.

For every fit we obtain a value for the reduced $\chi^2$, and $\chi^2_\nu=\chi^2/\nu$, where $\nu$ is the number of
free parameters, i.e. the number of observational data points minus the number of parameters used in the theoretical
model. For a good fit, $\chi^2_\nu$ should be about unity. The fit with the minimum value of $\chi^2_\nu$ [hereafter
$\chi^2_\nu (\rm {min})$] was adopted as the best fit and age was adopted. This method was applied for fits with the
BC03 models. To estimate the uncertainty in the determined age we use confidence limits.
If $\chi^2_\nu<\chi^2_\nu (\rm {min})+1$ then the resulting age is within the 68.3 per cent probability range. So the accepted
range in age is derived from the fits which have $\chi^2_\nu (\rm {min})<\chi^2_\nu<\chi^2_\nu (\rm {min})+1$.
With this method we derived the ages with their uncertainties of GC-1 and GC-2. For GC-1, the best-reduced
$\chi^2_\nu (\rm {min})=0.7$ is achieved with an age of $15.50\pm3.20$ Gyr when the value of $E(B-V)$ is adopted to
be 0.09, and $\nu=16$. For GC-2, the best-reduced $\chi^2_\nu (\rm {min})=2.9$ is achieved with an age of
$15.10\pm2.70$ Gyr when the value of $E(B-V)$ is adopted to be 0.10, and $\nu=16$. The values for the extinction
coefficient, $R_\lambda$, are obtained by interpolating the interstellar extinction curve of \citet{car89}. Fig. 2 shows the intrinsic SEDs of
GC-1 and GC-2, the integrated SEDs of the best-fitting models and the spectra of the best-fitting models. From Fig. 2,
we can see that, for GC-2, the BC03 SSP models cannot fit the photometric datum in the $c$ band as well as the other 17
photometric data, i.e. the observed magnitude is dimmer than that of the model in the $c$ band. We have not been able to
identify the cause of this discrepancy. Direct inspection of the image in the $c$ band clearly shows that GC-2 is not unusual.
Also, we used the same image to determine the photometries for GC-1 and GC-2 at the same time. So, we argued that this
discrepancy may be from the intrinsic property of GC-2, i.e. it is caused by the intrinsic luminosity of stellar populations
of GC-2. In the fitting, we did not use the magnitude in the $c$ band. In addition, it is evident that the ages  of GC-1 and GC-2
obtained here based on the BC03 models are greater than the currently accepted age of the universe. However, we must keep in mind
that the BC03 SSP models were calculated for ages up to 20 Gyr. In fact, ages derived for objects such as GCs and galaxies in
excess of that of the universe only mean that these objects are among the oldest objects in the universe.

\begin{figure*}
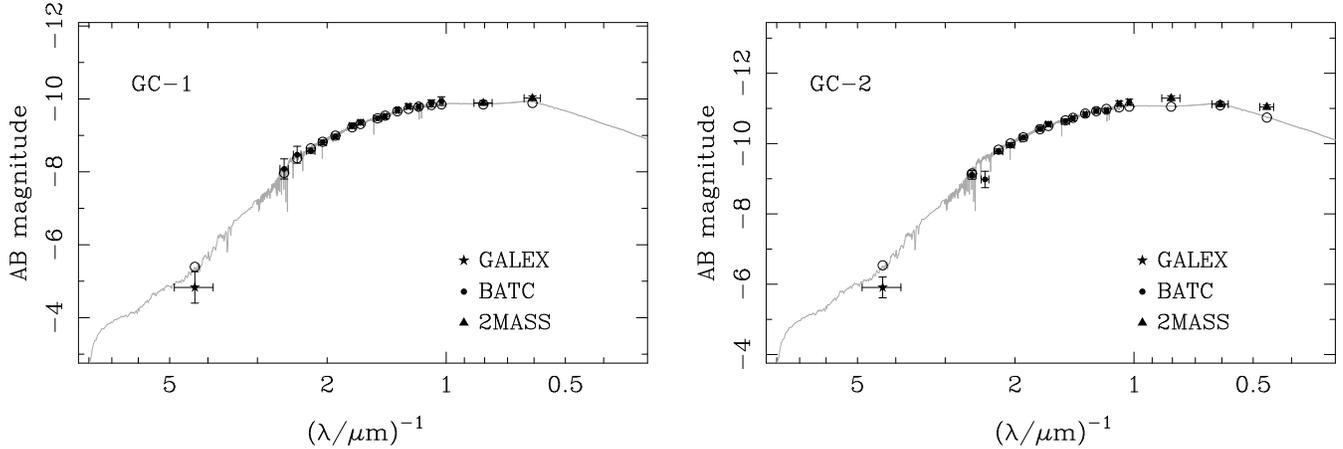

\centerline{\hfil
   \includegraphics[width=85mm]{fig2a.eps}
   \hfill
   \includegraphics[width=85mm]{fig2b.eps}
\hfil}
\caption{Best-fitting, integrated theoretical BC03 SEDs compared to the intrinsic SEDs of GC-1 and GC-2. The photometric
measurements are shown as symbols with error bars (vertical error bars for photometric uncertainties and horizontal ones
for the approximate wavelength coverage of each filter). The open circles represent the calculated magnitudes of the model
SED for each filter.}
\end{figure*}

\subsection{Masses}

We next determine the masses of GC-1 and GC-2. The BC03 SSP models are normalized to a total mass of 1 $\rm M_\odot$ in stars
at age $t=0$. The absolute magnitudes (in the Vega system) in $UBVRI$ and 2MASS $JHK_{\rm s}$ filters are included in the
BC03 SSP models. The difference between the intrinsic absolute magnitudes and those given by the model provides a direct
measurement of the cluster mass. To reduce mass uncertainties resulting from photometric uncertainties based on only
magnitudes in one filter (in general the $V$ band is used), we estimated the masses of GC-1 and GC-2 using magnitudes in
the $UBVRI$ and $JHK_{\rm s}$ bands. \citet{zhou03} have derived the relationships between the BATC intermediate-band system
and the broad-band system. However, we cannot derive the magnitudes in the $U$ band since the magnitudes in the $a$ band cannot
be obtained in this paper. The magnitudes in the $BVRI$ bands are obtained using the photometries of the BATC intermediate-band
system obtained here (see Table 2 for details) and the relationships between these two systems in \citet{zhou03}. The $BVRI$
magnitudes are $B=19.42\pm0.083$, $V=18.71\pm0.032$, $R=18.43\pm0.009$ and $I=17.73\pm0.025$ for GC-1, and $B=18.60\pm0.039$,
$V=17.79\pm0.016$, $R=17.51\pm0.005$ and $I=16.78\pm0.001$ for GC-2. The resulting mass determinations for GC-1 and GC-2 are
listed in Table 3 with their $1\sigma$ uncertainties. Table 3 shows that the masses of GC-1 and GC-2 obtained based on the
magnitudes in different bands are consistent.

\begin{table*}
\begin{minipage}{170mm}
\scriptsize
\caption{Mass estimates of GC-1 and GC-2 based on the BC03 models.} \label{table3}
\begin{tabular}{cccccccc}
\hline
  & $B$ & $V$ & $R$ & $I$  &  $J$ & $H$ & $K_{\rm s}$     \\
\hline
\multicolumn{8}{c}{Mass $(10^6~\rm M_\odot)$} \\
\hline
 GC-1   & $1.93 \pm 0.15 $ & $1.97 \pm 0.06 $ & $1.77  \pm 0.01 $ & $2.04 \pm0.05 $ & $1.87 \pm 0.05 $ & $2.04 \pm 0.04 $ & ...               \\
 GC-2   & $5.20 \pm 0.19 $ & $5.81 \pm 0.09 $ & $5.25  \pm 0.02 $ & $6.08 \pm0.01 $ & $6.78 \pm 0.11 $ & $5.61 \pm 0.07 $ & $7.11  \pm 0.08 $ \\
\hline
\end{tabular}
\end{minipage}
\end{table*}


\section{$HST$ data analysis}

\subsection{Ellipticity, position angle and surface brightness profile}
\label{brightness.sec}

The imaging data that we used in this work to obtain the structural and dynamical parameters of GC-1 and GC-2 come from the
{\it HST} observations in the $F606W$ and $F814W$ bands (see Section 2.4 for details). The data analysis procedures used to measure
the surface brightness profiles of clusters have been described in \citet{barmby07}; here we briefly summarize the procedures.
The surface photometries of each cluster were obtained from the drizzled images using the {\sc iraf} task {\sc ellipse}. The
centre position of each cluster was fixed at a value derived by the object locator of the {\sc ellipse} task; however an initial
centre position was determined by centroiding. Elliptical isophotes were fitted to the data, with no sigma clipping. We ran two
passes of the {\sc ellipse} task: the first pass was run in the usual way, i.e. the ellipticity and position angle (PA) vary with the
isophote semimajor axis. In the second pass the surface brightness profiles on fixed, zero-ellipticity isophotes were measured,
since we chose to fit circular models for the intrinsic cluster structure and the point spread function (PSF) as in \citet{barmby07}.
Fig. 3 and 4 plot the ellipticity ($\epsilon=1-b/a$) and PA as a function of the semimajor axis length ($a$) in
the $F606W$ and $F814W$ bands for GC-1 and GC-2. The errors were generated by the {\sc iraf} task {\sc ellipse}, in which the ellipticity
errors were obtained from the internal errors in the harmonic fit, after removal of the first and second fitted harmonics. The
ellipticities and PAs measured on images in different filters track well together, as we would expect. Figs 3 and 4
show that the PAs are occasionally wildly varying. This is likely to be produced by internal errors in the {\sc ellipse}
task. The final ellipticity and PA for each cluster were calculated as the average of the values obtained in the first
pass of {\sc ellipse} task, which are listed in Table 4. Errors correspond to the standard deviation of the mean. From Figs 3 and 4,
and Table 4, we can see that the ellipticities of GC-1 and GC-2 are less than 0.1, which are consistent with the mean value of
$0.07\pm0.01$ for the Milky Way GCs \citep{white87} and the mean $\epsilon=0.11\pm0.01$ for the M31 GCs \citep{bk02}.

\begin{table*}
\begin{minipage}{170mm}
\scriptsize
\caption{Basic parameters of GC-1 and GC-2 in the M81 group.} \label{table4}
\begin{tabular}{cccccccc}
\hline
ID & $\epsilon^{a}$ & $\theta^{a}$ & $\epsilon^{b}$ & $\theta^{b}$ & $E(B-V)^{c}$  & ${\rm [Fe/H]}^{d}$  & ${\rm Age}^{e}$ \\
    &      &  (deg E of N)  &  & (deg E of N)  &   &  &  (Gyr)  \\
(1) &  (2) &  (3)     &   (4)  & (5)  &  (6)  &  (7) &  (8)     \\
\hline
 GC-1   & $0.10 \pm 0.01 $ & $60   \pm 1    $ & $0.07  \pm 0.02 $ & $80   \pm5    $ & 0.09  & $-2.23 \pm 0.11 $ & $15.50 \pm 3.20 $ \\
 GC-2   & $0.08 \pm 0.01 $ & $-92  \pm 10   $ & $0.09  \pm 0.01 $ & $-106 \pm1    $ & 0.10  & $-2.30 \pm 0.12 $ & $15.10 \pm 2.70 $ \\
\hline
\end{tabular}

\medskip
$^{a}$ $\epsilon$ and $\theta$ are ellipticity and PA of $F606W$ filter for each cluster, respectively, which were obtained in this paper.

$^{b}$ $\epsilon$ and $\theta$ are ellipticity and PA of $F814W$ filter for each cluster, respectively, which were obtained in this paper.

$^{c}$ $E(B-V)$ is the reddening value of each cluster adopted in this paper.

$^{d}$ ${\rm [Fe/H]}$ is the metallicity of each cluster adopted in this paper.

$^{e}$ ${\rm Age}$ is the age of each cluster obtained in this paper.

\end{minipage}
\end{table*}


\begin{figure}
\resizebox{\hsize}{!}{\rotatebox{-90}{\includegraphics{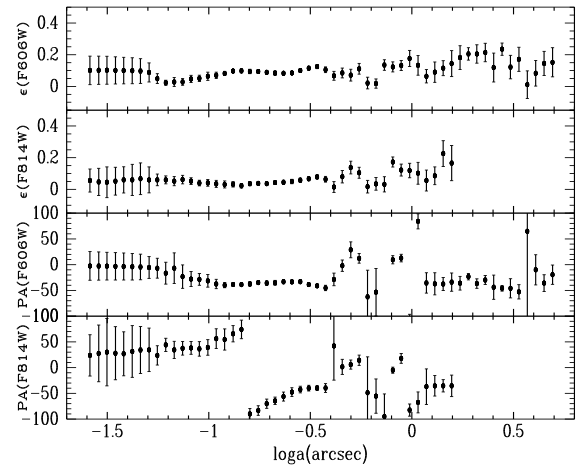}}}
\caption{Ellipticity ($\epsilon$) and PA as a function of the semimajor axis ($a$) in the $F606W$ and $F814W$
filters of {\it HST}/ACS/WFC for GC-1.}
\label{fig3}
\end{figure}

\begin{figure}
\resizebox{\hsize}{!}{\rotatebox{-90}{\includegraphics{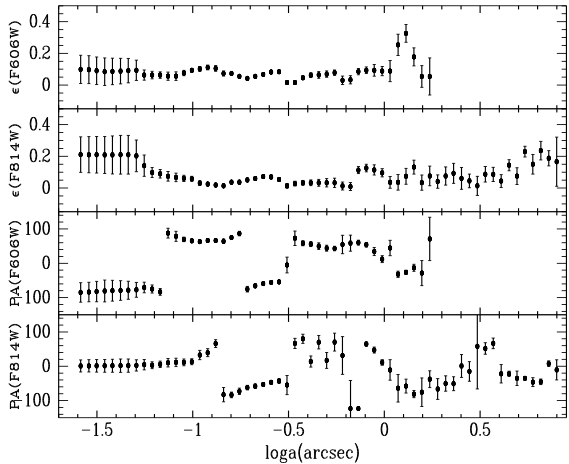}}}
\caption{Ellipticity ($\epsilon$) and PA as a function of the semimajor axis ($a$) in the $F606W$ and $F814W$
filters of {\it HST}/WFC3/UVIS for GC-2.}
\label{fig4}
\end{figure}

Raw output from package {\sc ellipse} is in terms of counts s$^{-1}$ pixel$^{-1}$, which needs to multiply by a number (400
for ACS/WFC, and 625 for WFC3/UVIS) to convert to counts s$^{-1}$ arcsec$^{-2}$. A formula is used to transform the counts
to surface brightness in magnitudes calibrated on the {\sc vegamag} system,

\begin{equation}
\mu/{\rm mag~arcsec^{-2}=
-2.5 \log(counts~s^{-1}~arcsec^{-2}) + Zeropoint}.
\end{equation}

As noted by \citet{barmby07}, occasional oversubtraction of the background during the multidrizzling in the automatic reduction
pipeline leads to ``negative'' counts in some pixels, so we worked in terms of linear intensity instead of surface brightness in
magnitudes as \citet{barmby07} did. With updated absolute magnitudes of the Sun, ${\rm M}_{\odot}$ (C. Willmer, private communication)
listed in Table 5, the equation for transforming counts to surface brightness in intensity is derived as below,

\begin{table*}
\begin{minipage}{170mm}
\scriptsize
\caption{Calibration data for the {\it HST} images.} \label{table5}
\begin{tabular}{ccccccc}
\hline
Filter & Pivot $\lambda$ & $R_{\lambda}$${^a}$ & Zeropoint${^b}$ & ${\rm M}_{\odot}$${^c}$  & Conversion Factor${^d}$  & Coefficient${^e}$ \\
       &   (\AA)         &    &   &  &  &   \\
(1) &  (2)  &  (3) &  (4) & (5) &  (6)  & (7)\\
\multicolumn{7}{c}{Calibration Data for ACS/WFC images} \\
\hline
$F606W$  &  5921.1 & 2.85   & 26.398  &  4.611     &  0.8207     &   26.183     \\
$F814W$  &  8057.0 & 1.83   & 25.501  &  4.066     &  1.1349     &   25.638     \\
\hline
\multicolumn{7}{c}{Calibration Data for WFC3/UVIS images} \\
\hline
$F606W$  &  5887.0 & 2.87   & 25.987  &  4.640     &   1.2312    &   26.212   \\
$F814W$  &  8024.0 & 1.84   & 24.680  &  4.100     &   2.4937    &   25.672   \\
\hline
\end{tabular}

\medskip
$^{a}$ $A_{\lambda}=R_{\lambda} \times E(B-V)$.

$^{b}$ Additive conversion between surface brightness in counts s$^{-1}$ arcsec$^{-2}$ and magnitude in mag arcsec$^{-2}$.

$^{c}$ Updated absolute magnitude of the Sun (C. Willmer, private communication).

$^{d}$ Multiplicative conversion between surface brightness in counts s$^{-1}$ arcsec$^{-2}$ and intensity in ${\rm L}_{\odot}$ pc$^{-2}$.

$^{e}$ Additive conversion between surface brightness in magnitude in mag arcsec$^{-2}$ and intensity in ${\rm L}_{\odot}$ pc$^{-2}$.

\end{minipage}
\end{table*}


\begin{equation}
I/L_{\odot}~{\rm pc^{-2}
\simeq Conversion~Factor\times(counts~s^{-1}~arcsec^{-2})}.
\end{equation}

Converting from luminosity density in ${\rm L}_{\odot}~{\rm pc^{-2}}$ to surface brightness in magnitude was done according to
\begin{equation}
\mu/{\rm mag~arcsec^{-2}}=
-2.5\log(I/{\rm L}_{\odot}~{\rm pc^{-2}}) + {\rm Coefficient}.
\end{equation}

The zero-points, conversion factors and coefficients used in these transformations for each filter are also listed in Table 5.
In this paper, the final, calibrated intensity profiles for GC-1 and GC-2 with no extinction correction are listed in Table 6.
The reported intensities are calibrated on the {\sc vegamag} scale. In Table 6, Column 7 gives a flag for each point, which has
the same meaning as defined by \citet{barmby07} and \citet{mclau08}. The points flagged with `OK' are used to constrain the model
fit, while the points flagged with `DEP' are those that may lead to excessive weighting of the central regions of clusters
\citep[see][for details]{barmby07,mclau08}. In addition, points marked with `BAD' are those individual isophotes that deviated
strongly from their neighbors or showed irregular features, which were deleted by hand.

\begin{table}
\begin{minipage}{170mm}
\scriptsize
\caption{Intensity profiles for GC-1 and GC-2 in the M81 group.} \label{table6}
\begin{tabular}{cccccccc}
\hline
ID & Detector & Filter & $R^{a}$ & $I$ & Uncertainty  & Flag   \\
     &       &    & (arcsec)  &  (${\rm L}_{\odot}$~pc$^{-2}$)   & (${\rm L}_{\odot}$~pc$^{-2}$)  &   \\
(1) &  (2)  &  (3) &  (4) & (5) &  (6)  & (7)\\
\hline
 GC-1     & ACS/WFC    & $F606W$    & 0.0260     & 8027.168     & 152.402      & OK    \\
          & ACS/WFC    & $F606W$    & 0.0287     & 7939.514     & 166.932      & DEP   \\
          & ACS/WFC    & $F606W$    & 0.0315     & 7843.411     & 182.646      & DEP   \\
          & ACS/WFC    & $F606W$    & 0.0347     & 7738.249     & 199.672      & DEP   \\
          & ACS/WFC    & $F606W$    & 0.0381     & 7623.505     & 218.122      & DEP   \\
          & ACS/WFC    & $F606W$    & 0.0420     & 7498.559     & 238.123      & DEP   \\
          & ACS/WFC    & $F606W$    & 0.0461     & 7362.532     & 259.738      & DEP   \\
          & ACS/WFC    & $F606W$    & 0.0508     & 7211.208     & 282.077      & DEP   \\
          & ACS/WFC    & $F606W$    & 0.0558     & 6985.240     & 292.158      & OK    \\
          & ACS/WFC    & $F606W$    & 0.0614     & 6723.962     & 300.276      & DEP   \\
          & ACS/WFC    & $F606W$    & 0.0676     & 6413.738     & 305.292      & DEP   \\
          & ACS/WFC    & $F606W$    & 0.0743     & 6069.431     & 303.803      & DEP   \\
          & ACS/WFC    & $F606W$    & 0.0818     & 5690.524     & 307.635      & OK    \\
          & ACS/WFC    & $F606W$    & 0.0899     & 5265.273     & 304.361      & DEP   \\
          & ACS/WFC    & $F606W$    & 0.0989     & 4809.377     & 306.135      & DEP   \\
          & ACS/WFC    & $F606W$    & 0.1088     & 4313.025     & 273.523      & DEP   \\
          & ACS/WFC    & $F606W$    & 0.1197     & 3832.797     & 238.625      & DEP   \\
          & ACS/WFC    & $F606W$    & 0.1317     & 3417.995     & 209.960      & OK    \\
          & ACS/WFC    & $F606W$    & 0.1448     & 2942.204     & 178.856      & OK    \\
          & ACS/WFC    & $F606W$    & 0.1593     & 2425.916     & 142.266      & OK    \\
\hline
\end{tabular}

\medskip
$^{a}$ $R$ is the clustercentric radius.

$Note.$ This table is available in its entirely in machine-readable form.

\end{minipage}
\end{table}


\subsection{Point-spread function}

The PSF models are critical to accurately measure the shapes of the objects in images taken with {\it HST}
\citep{rhodes06}. In this paper, we chose not to deconvolve the data, instead fitting the structural models after convolving them
with a simple analytic description of the PSF as \citet{barmby07} and \citet{wang13} did \citep[see][for details]{barmby07,wang13}
for M31 star clusters. A simple analytic description of the PSFs for the ACS/WFC filters has been given by \citet{wang13}. In addition,
we derived the WFC3/UVIS PSF models using Tiny Tim \footnote{http://tinytim.stsci.edu/cgi-bin/tinytimweb.cgi.} as \citet{wang13} did
[see equation (4) of Wang \& Ma 2013]. Table 7 lists the parameters. The full width at half-maximum (FWHMs) of the adopted PSF in each filter are:
0.067 arcsec for the ACS/WFC $F606W$, 0.071 arcsec for the ACS/WFC $F814W$, 0.069 arcsec for the WFC3/UVIS $F606W$ and 0.080 arcsec for
the WFC3/UVIS $F814W$.

\begin{table}
\begin{minipage}{170mm}
\scriptsize
\caption{Coefficients for the PSF models.} \label{table7}
\begin{tabular}{lcccc}
\hline
Detector & Filter & $r_0$ & $\alpha$ & $\beta$  \\
    &       & (arcsec)   &   &  \\
(1) &  (2)  &  (3) &  (4) & (5) \\
\hline
ACS/WFC   &  $F606W$ &  0.053    &   3    &   3.14     \\
          &  $F814W$ &  0.056    &   3    &   3.05     \\
WFC3/UVIS &  $F606W$ &  0.050    &   3    &   3.24     \\
          &  $F814W$ &  0.058    &   3    &   3.50     \\
\hline
\end{tabular}
\end{minipage}
\end{table}


\subsection{Extinction and magnitude transformation}

When we fit models to the surface brightness profiles of GC-1 and GC-2, we correct the intensity profiles for extinction.
Table 5 lists the effective wavelengths of the ACS/WFC and WFC3/UVIS filters from the Instrument Handbook. With the extinction
curve taken from \citet{car89} with $R_V=3.1$, we derived the $A_{\lambda}$ values for each filter. The reddening values of GC-1
and GC-2 having mentioned and used in Section 3 are listed in Table 4.

\section{Model fitting}
\label{fit.sec}

\subsection{Structural models}

As was done by \citet{barmby07}, \citet{mclau08}, \citet{wang13}, and \citet{ma15}, we used three structural models to fit surface
profiles of GC-1 and GC-2. These models are developed by \citet{king66}, \citet{wilson75} and \citet{sersic68} (hereafter
the `King model', `Wilson model' and `S\'{e}rsic model', respectively). \citet{mclau08} have described the three structural
models in detail; here we briefly summarize some of their basic characteristics.

The King model is most commonly used for studies of star clusters, which is the simple model of single-mass, isotropic, modified
isothermal sphere. The Wilson model is defined by an alternate modified isothermal sphere based on the ad hoc stellar distribution
function of \citet{wilson75}, which has more extended envelope structures than the standard King isothermal sphere \citep{mclau08}.
The S\'{e}rsic model has an $R^{1/n}$ surface density profile, which is used for parameterizing the surface brightness profiles of
early-type galaxies and bulges of spiral galaxies \citep{bg11}. However, \citet{tanvir12} found that some classical GCs in M31 that
exhibit cuspy core profiles are well fitted by the S\'{e}rsic model of index $n\sim2-6$. The clusters with cuspy cores have usually
been called post-core collapse \citep[see][and references therein]{ng06}.

\subsection{Fits}

After the intensity profiles were corrected for extinction (see Section 4.3 for details), we fitted models to the brightness profiles
of GC-1 and GC-2.

We first convolved the three models with PSFs for the filters used. Given a value for the scale radius $r_0$, we computed a dimensionless
model profile $\widetilde{I}_{\rm mod}\equiv I_{\rm mod}/I_0$ and then performed the convolution,

\begin{equation}
\widetilde{I}_{\rm mod}^{*} (R | r_0) = \int\!\!\!\int_{-\infty}^{\infty}
               \widetilde{I}_{\rm mod}(R^\prime/r_0)
               \widetilde{I}_{\rm PSF}
               \left[(x-x^\prime),(y-y^\prime)\right]{\rm d}x^\prime {\rm d}y^\prime,
\label{eq:convol}
\end{equation}
where $R^2=x^2+y^2$ and $R^{\prime2}=x^{\prime2}+y^{\prime2}$. $\widetilde{I}_{\rm PSF}$ was approximated using equation 4 of
\citet{wang13} \citep[see][for details]{mclau08}. The best-fitting model was derived by calculating and minimizing $\chi^2$ as the
sum of squared differences between model intensities and observed intensities with the extinction corrected,
\begin{equation}
\chi^2=\sum_{i}{\frac{[I_{\rm obs}(R_i)-I_0\widetilde{I}_{\rm mod}^{*}(R_i|r_0)
       -I_{\rm bkg}]^2}{\sigma_{i}^{2}}},
\end{equation}
in which a background $I_{\rm bkg}$ was also fitted. The uncertainties of observed intensities listed in Table 6 were used as weights.

We plot the fitting for GC-1 and GC-2 in Figs 5 and 6. The observed intensity profile with extinction corrected is plotted as a
function of logarithmic projected radius. The open squares are the data points included in the model fitting, while the crosses are
points flagged as `DEP' or `BAD', which are not used to constrain the fit \citep{wang13}. The best-fitting models, including the
King model (K66), Wilson model (W) and S\'{e}rsic model (S), are shown with a red solid line from the left-hand to the right-hand panel, with
a fitted $I_{\rm bkg}$ added. The blue dashed lines represent the shapes of the PSFs for the filters used.

\begin{figure}
\centerline{\hfil
   \includegraphics[width=82mm]{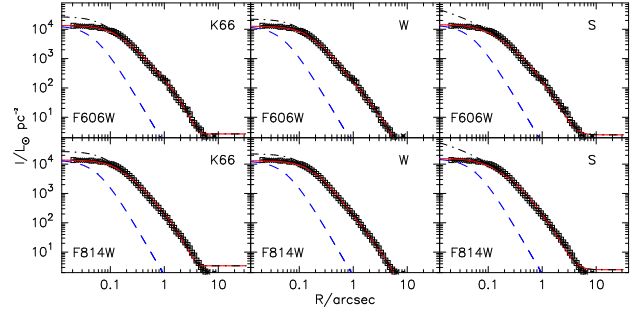}
\hfil}
\caption{Surface brightness profiles of and model fits to GC-1, with the data of the $F606W$ and $F814W$ bands from top to bottom. The
three panels in each line from left to right are the fits to the King model (K66), Wilson model (W) and S\'{e}rsic model (S). The
open squares are the data points included in the model fitting, while the crosses are points flagged as `DEP' or `BAD', which
are not used to constrain the fit. Dashed (blue) curves trace the PSF intensity profile; bold dash-dotted curves,
the unconvolved best-fitting model and solid (red) curves, the PSF-convolved best fits.
\label{fig:fig5}
}
\end{figure}

\begin{figure}
\centerline{\hfil
   \includegraphics[width=82mm]{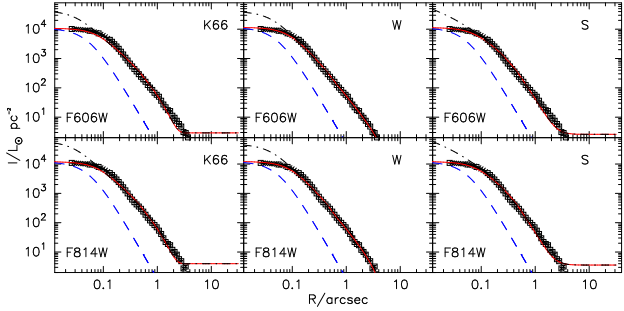}
\hfil}
\caption{Same as Fig. 5, but with surface brightness profiles of and model fits to GC-2.
\label{fig:fig6}
}
\end{figure}

\subsection{Main model parameters and derived qualities}

Table 8 lists the basic ingredients of all model fits to GC-1 and GC-2, with a simple description of each parameter/column at
the end. Error bars on all these parameters (and those on all the derived quantities below) were defined in the same
way as in \citet{mclau08}.

\begin{table*}
\begin{minipage}{170mm}
\scriptsize
\caption{Basic parameters of GC-1 and GC-2 in the M81 group.}\label{table8}
\begin{tabular}{ccccccrcccrr}
\hline
ID & Detector & Band & $N_{\rm pts}$${^a}$ & Model & $\chi_{\rm min}^2$${^b}$  & $I_{\rm bkg}$${^c}$  & $W_0$${^d}$  &
$c/n$${^e}$ & $\mu_0$${^f}$  & $\log r_0$${^g}$   &  $\log r_0$${^h}$  \\
     &       &    &   &     &   & (${\rm L}_{\odot}$~pc$^{-2}$)  &  &   &  ${\rm (mag~arcsec^{-2})}$ & (arcsec) & (pc)  \\
(1) &  (2)  &  (3) &  (4) & (5) &  (6)  & (7)  &  (8)   &  (9)   &   (10)   &   (11)  &   (12)  \\
\hline
 GC-1       & ACS/WFC    & 606       & 40       & K66      & 48.87    & $2.90     \pm 1.20    $ & $8.50    ^{+0.60    }_{-0.57   }         $ & $1.98    ^{+0.16    }_{-0.17   }         $ & $16.11   ^{+0.25    }_{-0.15   }$ & $-1.450  ^{+0.015   }_{-0.017  }$ & $-0.205  ^{+0.015   }_{-0.017  }$\\
          &            &           & 40       & W        & 8.84     & $-0.40    \pm 1.15    $ & $7.90    ^{+0.29    }_{-0.27   }         $ & $3.08    ^{+0.14    }_{-0.21   }         $ & $16.03   ^{+0.07    }_{-0.06   }$ & $-1.350  ^{+0.011   }_{-0.013  }$ & $-0.105  ^{+0.011   }_{-0.013  }$\\
          &            &           & 40       & S        & 48.78    & $2.60     \pm 0.80    $ & $-                                  $ & $3.10    ^{+0.08    }_{-0.08   }         $ & $16.05   ^{+0.23    }_{-0.17   }$ & $-3.500  ^{+0.026   }_{-0.046  }$ & $-2.255  ^{+0.026   }_{-0.046  }$\\
        & ACS/WFC    & 814       & 38       & K66      & 32.07    & $4.00     \pm 1.98    $ & $8.60    ^{+0.57    }_{-0.51   }         $ & $2.01    ^{+0.15    }_{-0.15   }         $ & $15.45   ^{+0.17    }_{-0.14   }$ & $-1.500  ^{+0.031   }_{-0.065  }$ & $-0.255  ^{+0.031   }_{-0.065  }$\\
          &            &           & 38       & W        & 10.90    & $-1.20    \pm 2.55    $ & $7.90    ^{+0.45    }_{-0.36   }         $ & $3.08    ^{+0.18    }_{-0.28   }         $ & $15.43   ^{+0.08    }_{-0.08   }$ & $-1.350  ^{+0.014   }_{-0.015  }$ & $-0.105  ^{+0.014   }_{-0.015  }$\\
          &            &           & 38       & S        & 40.30    & $3.60     \pm 1.91    $ & $-                                  $ & $3.05    ^{+0.11    }_{-0.13   }         $ & $17.95   ^{+0.71    }_{-0.62   }$ & $-3.450  ^{+0.031   }_{-0.035  }$ & $-2.205  ^{+0.031   }_{-0.035  }$\\
 GC-2       & WFC3/UVIS  & 606       & 46       & K66      & 30.25    & $2.70     \pm 1.02    $ & $8.20    ^{+0.30    }_{-0.28   }         $ & $1.89    ^{+0.09    }_{-0.09   }         $ & $15.83   ^{+0.10    }_{-0.07   }$ & $-1.100  ^{+0.031   }_{-0.015  }$ & $0.193   ^{+0.031   }_{-0.015  }$\\
          &            &           & 46       & W        & 10.88    & $-1.00    \pm 0.99    $ & $7.70    ^{+0.17    }_{-0.20   }         $ & $2.95    ^{+0.11    }_{-0.17   }         $ & $15.90   ^{+0.08    }_{-0.04   }$ & $-1.000  ^{+0.010   }_{-0.008  }$ & $0.293   ^{+0.010   }_{-0.008  }$\\
          &            &           & 46       & S        & 64.13    & $2.50     \pm 1.33    $ & $-                                  $ & $2.90    ^{+0.10    }_{-0.07   }         $ & $15.74   ^{+0.15    }_{-0.09   }$ & $-2.950  ^{+0.027   }_{-0.039  }$ & $-1.657  ^{+0.027   }_{-0.039  }$\\
        & WFC3/UVIS  & 814       & 46       & K66      & 22.97    & $3.40     \pm 0.90    $ & $7.90    ^{+0.25    }_{-0.27   }         $ & $1.80    ^{+0.07    }_{-0.08   }         $ & $15.22   ^{+0.10    }_{-0.08   }$ & $-1.050  ^{+0.021   }_{-0.018  }$ & $0.243   ^{+0.021   }_{-0.018  }$\\
          &            &           & 46       & W        & 4.44     & $-0.30    \pm 0.78    $ & $7.40    ^{+0.17    }_{-0.10   }         $ & $2.69    ^{+0.14    }_{-0.08   }         $ & $15.29   ^{+0.03    }_{-0.02   }$ & $-0.950  ^{+0.005   }_{-0.008  }$ & $0.343   ^{+0.005   }_{-0.008  }$\\
          &            &           & 46       & S        & 44.23    & $2.50     \pm 1.23    $ & $-                                  $ & $2.90    ^{+0.07    }_{-0.06   }         $ & $18.15   ^{+0.52    }_{-0.48   }$ & $-2.950  ^{+0.030   }_{-0.024  }$ & $-1.657  ^{+0.030   }_{-0.024  }$\\
\hline
\end{tabular}

\medskip
$^{a}$ The number of points in the intensity profile that were used for constraining the model fits.

$^{b}$ The minimum $\chi^2$ obtained in the fits.

$^{c}$ The best fitted background intensity.

$^{d}$ The dimensionless central potential of the best-fitting model, defined as $W_0 \equiv -\phi(0)/\sigma_0^2$.

$^{e}$ The concentration $c \equiv \log(r_t/r_0)$.

$^{f}$ The best-fitting central surface brightness in the native bandpass of the data.

$^{g}$ The best model fit scale radius $r_0$ in arcsec.

$^{h}$ The best model fit scale radius $r_0$ in parsec.

\end{minipage}
\end{table*}


Table 9 contains a number of other structural cluster properties derived from the basic fit parameters given in Table 8, with a
simple description of each parameter/column at the end, and the details of the calculations are presented in \citet{mclau08}.
The uncertainties on all of these derived parameters were estimated in the same way as in \citet{mclau08}.

\begin{table*}
\begin{minipage}{170mm}
\scriptsize
\caption{Derived structural and photometric parameters of GC-1 and GC-2 in the M81 group.}\label{table9}
\begin{tabular}{cccccrcccccccc}
\hline
ID & Detector & Band & Model & $\log r_{\rm t}$${^a}$  & $\log R_{\rm c}$${^b}$  & $\log R_{\rm h}$${^c}$  & $\log R_{\rm h}/R_{\rm c}$${^d}$ & $\log I_{\rm 0}$${^e}$
& $\log j_{\rm 0}$${^f}$   & $\log L_V$${^g}$ & $V_{\rm tot}$${^h}$ & $\log I_{\rm h}$${^i}$ & $\langle\mu_V\rangle_{\rm h}$${^j}$ \\
   &      &    &   &  (pc)   & (pc)  & (pc)  &   &  ($L_{\odot,V}$~pc$^{-2}$) & ($L_{\odot,V}$~pc$^{-3}$) &  ($L_{\odot,V}$) &  (mag) &  ($L_{\odot,V}$~pc$^{-2}$)
   &   ${\rm (mag~arcsec^{-2})}$  \\
(1) &  (2)  &  (3) &  (4) & (5) &  (6)  & (7)  &  (8)   &  (9)   &   (10)   &   (11)  &   (12)  &   (13)  &   (14)  \\
\hline
 GC-1       & ACS/WFC    & 606        & K66      & $1.78    ^{+0.17    }_{-0.17   }         $ & $-0.214  ^{+0.024   }_{-0.055  }$ & $0.740   ^{+0.196   }_{-0.144  }$ & $0.955   ^{+0.250   }_{-0.168  }$ & $4.02    ^{+0.06    }_{-0.10   }$ & $3.93    ^{+0.12    }_{-0.13   }$ & $5.63    ^{+0.04    }_{-0.03   }$ & $18.51   ^{+0.07    }_{-0.10   }$ & $3.35    ^{+0.26    }_{-0.35   }$ & $17.98   ^{+0.88    }_{-0.64   }$\\
          &            &            & W        & $2.98    ^{+0.14    }_{-0.18   }         $ & $-0.127  ^{+0.008   }_{-0.010  }$ & $0.876   ^{+0.146   }_{-0.166  }$ & $1.003   ^{+0.155   }_{-0.174  }$ & $4.05    ^{+0.03    }_{-0.03   }$ & $3.87    ^{+0.04    }_{-0.04   }$ & $5.74    ^{+0.04    }_{-0.05   }$ & $18.24   ^{+0.12    }_{-0.10   }$ & $3.19    ^{+0.28    }_{-0.25   }$ & $18.39   ^{+0.63    }_{-0.71   }$\\
          &            &            & S        & $\infty                                  $ & $-2.255  ^{+0.026   }_{-0.046  }$ & $0.699   ^{+0.171   }_{-0.149  }$ & $2.953   ^{+0.216   }_{-0.175  }$ & $4.05    ^{+0.07    }_{-0.10   }$ & $5.05    ^{+0.11    }_{-0.11   }$ & $5.64    ^{+0.08    }_{-0.07   }$ & $18.47   ^{+0.18    }_{-0.20   }$ & $3.45    ^{+0.23    }_{-0.26   }$ & $17.74   ^{+0.65    }_{-0.56   }$\\
        & ACS/WFC    & 814        & K66      & $1.76    ^{+0.15    }_{-0.15   }         $ & $-0.264  ^{+0.028   }_{-0.062  }$ & $0.736   ^{+0.188   }_{-0.140  }$ & $1.000   ^{+0.250   }_{-0.168  }$ &  & $3.98    ^{+0.13    }_{-0.13   }$ &  &  & $3.36    ^{+0.25    }_{-0.33   }$ & $17.96   ^{+0.84    }_{-0.63   }$\\
          &            &            & W        & $2.98    ^{+0.18    }_{-0.24   }         $ & $-0.127  ^{+0.010   }_{-0.011  }$ & $0.882   ^{+0.238   }_{-0.194  }$ & $1.009   ^{+0.249   }_{-0.204  }$ &  & $3.87    ^{+0.04    }_{-0.05   }$ &  &  & $3.18    ^{+0.34    }_{-0.44   }$ & $18.41   ^{+1.09    }_{-0.85   }$\\
          &            &            & S        & $\infty                                  $ & $-2.205  ^{+0.031   }_{-0.035  }$ & $0.693   ^{+0.164   }_{-0.142  }$ & $2.898   ^{+0.198   }_{-0.173  }$ &  & $5.01    ^{+0.09    }_{-0.11   }$ &  &  & $3.46    ^{+0.21    }_{-0.25   }$ & $17.71   ^{+0.62    }_{-0.53   }$\\
 GC-2       & WFC3/UVIS  & 606        & K66      & $2.09    ^{+0.09    }_{-0.09   }         $ & $0.181   ^{+0.030   }_{-0.013  }$ & $0.996   ^{+0.092   }_{-0.082  }$ & $0.815   ^{+0.105   }_{-0.112  }$ & $4.13    ^{+0.03    }_{-0.04   }$ & $3.65    ^{+0.05    }_{-0.07   }$ & $6.16    ^{+0.03    }_{-0.03   }$ & $17.42   ^{+0.07    }_{-0.07   }$ & $3.37    ^{+0.14    }_{-0.15   }$ & $17.93   ^{+0.39    }_{-0.34   }$\\
          &            &            & W        & $3.24    ^{+0.11    }_{-0.15   }         $ & $0.268   ^{+0.007   }_{-0.006  }$ & $1.130   ^{+0.100   }_{-0.132  }$ & $0.862   ^{+0.106   }_{-0.139  }$ & $4.11    ^{+0.03    }_{-0.04   }$ & $3.53    ^{+0.03    }_{-0.05   }$ & $6.25    ^{+0.03    }_{-0.04   }$ & $17.21   ^{+0.11    }_{-0.09   }$ & $3.19    ^{+0.22    }_{-0.17   }$ & $18.39   ^{+0.41    }_{-0.55   }$\\
          &            &            & S        & $\infty                                  $ & $-1.657  ^{+0.027   }_{-0.039  }$ & $0.970   ^{+0.126   }_{-0.119  }$ & $2.627   ^{+0.166   }_{-0.146  }$ & $4.17    ^{+0.04    }_{-0.06   }$ & $4.61    ^{+0.07    }_{-0.08   }$ & $6.15    ^{+0.08    }_{-0.07   }$ & $17.46   ^{+0.17    }_{-0.19   }$ & $3.41    ^{+0.17    }_{-0.18   }$ & $17.84   ^{+0.44    }_{-0.42   }$\\
        & WFC3/UVIS  & 814        & K66      & $2.05    ^{+0.08    }_{-0.08   }         $ & $0.229   ^{+0.019   }_{-0.016  }$ & $0.957   ^{+0.115   }_{-0.075  }$ & $0.727   ^{+0.132   }_{-0.094  }$ &  & $3.60    ^{+0.05    }_{-0.06   }$  &  &   & $3.45    ^{+0.12    }_{-0.20   }$ & $17.74   ^{+0.51    }_{-0.30   }$\\
          &            &            & W        & $3.04    ^{+0.14    }_{-0.07   }         $ & $0.315   ^{+0.004   }_{-0.006  }$ & $1.042   ^{+0.087   }_{-0.071  }$ & $0.727   ^{+0.093   }_{-0.075  }$ &  & $3.49    ^{+0.03    }_{-0.04   }$ &  &  & $3.37    ^{+0.10    }_{-0.14   }$ & $17.95   ^{+0.35    }_{-0.25   }$\\
          &            &            & S        & $\infty                                  $ & $-1.657  ^{+0.030   }_{-0.024  }$ & $0.969   ^{+0.127   }_{-0.120  }$ & $2.626   ^{+0.151   }_{-0.150  }$ &  & $4.61    ^{+0.06    }_{-0.08   }$ &  &  & $3.41    ^{+0.17    }_{-0.18   }$ & $17.83   ^{+0.45    }_{-0.43   }$\\
\hline
\end{tabular}

\medskip
$^{a}$ The model tidal radius $r_{\rm t}$ in parsec.

$^{b}$ The projected core radius of the model fitting a cluster, defined as $I(R_{\rm c}) = I_0/2$.

$^{c}$ The projected half-light, or effective radius of a model, containing half the total luminosity in projection.

$^{d}$ A measure of cluster concentration and relatively more model-independent than $W_0$ or $c$.

$^{e}$ The best-fitting central ($R = 0$) luminosity surface density in the $V$ band, defined as $\log I_0 = 0.4(26.358 - \mu_{V,0}$),
where 26.358 is the ``coefficient'' corresponding to a solar absolute magnitude $M_{V,\odot} = +4.786$ (C. Willmer, private communication).

$^{f}$ The central ($r = 0$) luminosity volume density in the $V$ band.

$^{g}$ The $V$-band total integrated model luminosity.

$^{h}$ The total $V$-band magnitude of a model cluster, defined as $V_{\rm tot} = 4.786 - 2.5 \log (L_V/{\rm L}_{\odot}) + 5\log (D/10$ pc).

$^{i}$ The luminosity surface density averaged over the half-light/effective radius in the $V$ band, defined as $\log I_{\rm h} \equiv \log (L_V/2{\pi}R_{\rm h}^2$).

$^{j}$ The surface brightness in magnitude over the half-light/effective radius in the $V$ band, defined as $\langle\mu_V\rangle_{\rm h} = 26.358 - 2.5\log I_{\rm h}$.
\end{minipage}
\end{table*}


Table 10 lists a number of `dynamical' cluster properties derived from the structural parameters already given, plus a mass-to-light
ratio ($M/L$ values). As was done by \citet{wang13}, the values of the mass-to-light ratios ($\Upsilon_V^{\rm pop}$ in Table 10) in the
$V$ band, which were used to derive the `dynamical' parameters, were determined from the population-synthesis models of BC03,
assuming a \citet{chab03} IMF. The ages and metallicities used to compute the $\Upsilon_V^{\rm pop}$ values in the $V$
band are listed in Table 4. The error bars on $\Upsilon_V^{\rm pop}$ include uncertainties in age and [Fe/H] listed in Table 4. The
uncertainties in these quantities are estimated from their variations around the minimum of $\chi^2$ on the model grids we fit combined
in quadrature with the uncertainties in the population-synthesis model $\Upsilon_V^{\rm pop}$.

\begin{table*}
\begin{minipage}{170mm}
\scriptsize
\caption{Derived dynamical parameters of GC-1 and GC-2 in the M81 group.}\label{table10}
\begin{tabular}{cccccccccccccc}
\hline
ID & Detector & Band & $\Upsilon_V^{\rm pop}$${^a}$ & Model  & $\log M_{\rm tot}$${^b}$  & $\log E_{\rm b}$${^c}$  & $\log \Sigma_{\rm 0}$${^d}$
 & $\log \rho_{\rm 0}$${^e}$  & $\log \Sigma_{\rm h}$${^f}$   & $\log \sigma_{p,0}$${^g}$ & $\log \nu_{\rm esc,0}$${^h}$ & $\log t_{\rm r,h}$${^i}$
 & $\log f_{\rm 0}$${^j}$ \\
 &   &   & (${\rm M}_{\odot}~{\rm L}_{\odot,V}^{-1}$)  &   & (${\rm M}_{\odot}$)  &  (erg)  &  (${\rm M}_{\odot}$~pc$^{-2}$)  &  (${\rm M}_{\odot}$~pc$^{-3}$) & (${\rm M}_{\odot}$~pc$^{-2}$) &
 (km s$^{-1}$) &  (km s$^{-1}$) &  (yr) &  [${\rm M}_{\odot}$~(pc~km~s$^{-1})^{-3}$]  \\
(1) &  (2)  &  (3) &  (4) & (5) &  (6)  & (7)  &  (8)   &  (9)   &   (10)   &   (11)  &   (12)  &   (13)  &   (14)  \\
\hline
 GC-1       & ACS/WFC    & 606        & $1.935   ^{+0.238   }_{-0.247  }         $ & K66      & $5.92    ^{+0.06    }_{-0.07   }$ & $50.43   ^{+0.22    }_{-0.33   }$ & $4.31    ^{+0.08    }_{-0.12   }$ & $4.22    ^{+0.13    }_{-0.14   }$ & $3.64    ^{+0.35    }_{-0.26   }$ & $0.786   ^{+0.026   }_{-0.048  }$ & $1.409   ^{+0.031   }_{-0.059  }$ & $9.56    ^{+0.31    }_{-0.24   }$ & $0.648   ^{+0.112   }_{-0.033  }$\\
          &            &            & $                                        $ & W        & $6.03    ^{+0.06    }_{-0.08   }$ & $50.65   ^{+0.22    }_{-0.25   }$ & $4.34    ^{+0.06    }_{-0.07   }$ & $4.16    ^{+0.06    }_{-0.07   }$ & $3.47    ^{+0.26    }_{-0.29   }$ & $0.846   ^{+0.027   }_{-0.032  }$ & $1.465   ^{+0.029   }_{-0.035  }$ & $9.81    ^{+0.24    }_{-0.28   }$ & $0.398   ^{+0.028   }_{-0.030  }$\\
          &            &            & $                                        $ & S        & $5.93    ^{+0.10    }_{-0.09   }$ & $48.47   ^{+0.29    }_{-0.40   }$ & $4.33    ^{+0.09    }_{-0.11   }$ & $5.34    ^{+0.12    }_{-0.13   }$ & $3.74    ^{+0.27    }_{-0.23   }$ & $0.131   ^{+0.039   }_{-0.060  }$ & $0.969   ^{+0.028   }_{-0.046  }$ & $9.50    ^{+0.29    }_{-0.26   }$ & $4.510   ^{+0.070   }_{-0.030  }$\\
        & ACS/WFC    & 814        & $1.935   ^{+0.238   }_{-0.247  }         $ & K66      &  & $50.30   ^{+0.21    }_{-0.32   }$ &  & $4.27    ^{+0.14    }_{-0.14   }$ & $3.65    ^{+0.34    }_{-0.26   }$ & $0.761   ^{+0.025   }_{-0.047  }$ & $1.387   ^{+0.028   }_{-0.056  }$ & $9.55    ^{+0.30    }_{-0.23   }$ & $0.772   ^{+0.131   }_{-0.038  }$\\
          &            &            & $                                        $ & W        &  & $50.65   ^{+0.23    }_{-0.26   }$ &  & $4.16    ^{+0.07    }_{-0.07   }$ & $3.46    ^{+0.44    }_{-0.34   }$ & $0.846   ^{+0.027   }_{-0.032  }$ & $1.465   ^{+0.031   }_{-0.035  }$ & $9.81    ^{+0.38    }_{-0.32   }$ & $0.398   ^{+0.031   }_{-0.032  }$\\
          &            &            & $                                        $ & S        &  & $48.49   ^{+0.38    }_{-0.49   }$ &  & $5.29    ^{+0.10    }_{-0.12   }$ & $3.75    ^{+0.25    }_{-0.22   }$ & $0.145   ^{+0.048   }_{-0.066  }$ & $0.974   ^{+0.031   }_{-0.044  }$ & $9.49    ^{+0.28    }_{-0.25   }$ & $4.394   ^{+0.041   }_{-0.030  }$\\
 GC-2       & WFC3/UVIS  & 606        & $1.939   ^{+0.236   }_{-0.249  }         $ & K66      & $6.45    ^{+0.06    }_{-0.07   }$ & $51.79   ^{+0.21    }_{-0.24   }$ & $4.42    ^{+0.06    }_{-0.07   }$ & $3.94    ^{+0.07    }_{-0.09   }$ & $3.66    ^{+0.16    }_{-0.15   }$ & $1.040   ^{+0.027   }_{-0.031  }$ & $1.657   ^{+0.030   }_{-0.032  }$ & $10.17   ^{+0.16    }_{-0.15   }$ & $-0.401  ^{+0.031   }_{-0.063  }$\\
          &            &            & $                                        $ & W        & $6.53    ^{+0.06    }_{-0.07   }$ & $51.91   ^{+0.21    }_{-0.25   }$ & $4.40    ^{+0.06    }_{-0.07   }$ & $3.82    ^{+0.06    }_{-0.07   }$ & $3.48    ^{+0.17    }_{-0.23   }$ & $1.072   ^{+0.027   }_{-0.033  }$ & $1.687   ^{+0.028   }_{-0.035  }$ & $10.41   ^{+0.17    }_{-0.22   }$ & $-0.621  ^{+0.025   }_{-0.030  }$\\
          &            &            & $                                        $ & S        & $6.43    ^{+0.09    }_{-0.09   }$ & $50.01   ^{+0.30    }_{-0.32   }$ & $4.46    ^{+0.07    }_{-0.09   }$ & $4.89    ^{+0.08    }_{-0.10   }$ & $3.69    ^{+0.18    }_{-0.18   }$ & $0.451   ^{+0.033   }_{-0.044  }$ & $1.260   ^{+0.025   }_{-0.035  }$ & $10.12   ^{+0.23    }_{-0.22   }$ & $2.989   ^{+0.064   }_{-0.037  }$\\
        & WFC3/UVIS  & 814        & $1.939   ^{+0.236   }_{-0.249  }         $ & K66      &  & $51.89   ^{+0.21    }_{-0.25   }$ &  & $3.89    ^{+0.07    }_{-0.09   }$ & $3.74    ^{+0.21    }_{-0.13   }$ & $1.065   ^{+0.026   }_{-0.032  }$ & $1.675   ^{+0.028   }_{-0.035  }$ & $10.11   ^{+0.19    }_{-0.14   }$ & $-0.525  ^{+0.037   }_{-0.041  }$\\
          &            &            & $                                        $ & W        &  & $52.01   ^{+0.21    }_{-0.25   }$ &  & $3.77    ^{+0.06    }_{-0.07   }$ & $3.65    ^{+0.15    }_{-0.12   }$ & $1.095   ^{+0.027   }_{-0.034  }$ & $1.705   ^{+0.028   }_{-0.035  }$ & $10.27   ^{+0.15    }_{-0.14   }$ & $-0.741  ^{+0.025   }_{-0.031  }$\\
          &            &            & $                                        $ & S        &  & $50.01   ^{+0.27    }_{-0.31   }$ &  & $4.89    ^{+0.08    }_{-0.10   }$ & $3.70    ^{+0.19    }_{-0.18   }$ & $0.451   ^{+0.034   }_{-0.042  }$ & $1.260   ^{+0.027   }_{-0.034  }$ & $10.12   ^{+0.23    }_{-0.22   }$ & $2.989   ^{+0.037   }_{-0.042  }$\\
\hline
\end{tabular}

\medskip
$^{a}$ The $V$-band mass-to-light ratio.

$^{b}$ The integrated cluster mass, estimated as $\log M_{\rm tot} = \log \Upsilon_V^{\rm pop} + \log L_V$.

$^{c}$ The integrated binding energy in ergs, defined as $E_b \equiv -(1/2) \int_0^{r_{\rm t}} 4{\rm \pi}r^2\rho\phi{\rm d}r$ .

$^{d}$ The central surface mass density, estimated as $\log \Sigma_0=\log \Upsilon_V^{\rm pop} + \log I_0$.

$^{e}$ The central volume density, estimated as $\log \rho_0 = \log \Upsilon_V^{\rm pop} + \log j_0$.

$^{f}$ The surface mass density averaged over the half-light/effective radius $R_h$, estimated as $\log \Sigma_{\rm h} = \log \Upsilon_V^{\rm pop} + \log I_{\rm h}$.

$^{g}$ The predicted line-of-sight velocity dispersion at the cluster center.

$^{h}$ The predicted central `escape' velocity with which a star can move out from the center of a cluster, defined as
$\nu_{\rm esc,0}^2/\sigma_0^2 = 2[W_0 + GM_{\rm tot}/r_{\rm t}\sigma_0^2]$.

$^{i}$ The two-body relaxation time at the model-projected half-mass radius, estimated as
$t_{\rm r,h} =(2.06\times10^6 {\rm yr} /\ln(0.4M_{\rm tot}/m_{\star}))(M_{\rm tot}^{1/2}R_{\rm h}^{3/2}/m_{\star})$. $m_{\star}$
is the average stellar mass in a cluster, assumed to be 0.5 $\rm M_{\odot}$.

$^{j}$ The model's central phase-space density, defined as $\log f_0 \equiv \log [\rho_0/(2\pi\sigma_{\rm c}^2)^{3/2}$].
\end{minipage}
\end{table*}


We should emphasize that the `dynamical' parameters of GC-1 and GC-2 in Tables 9 and 10 are predicted values and are not directly
obtained from spectroscopy measurements. In addition, among the `dynamical' parameters listed in Tables 9 and 10, the predicted
line-of-sight velocity dispersions ($\sigma_{{\rm p},0}$) can be compared directly to spectroscopic determinations. However, it is
unfortunate that there are not spectroscopic data that can be used to derived measured velocity dispersions of GC-1 and GC-2.

At last, we calculated analogues of the so-called $\kappa$ parameters of GC-1 and GC-2, and listed them in Table 11. The `$\kappa$'
parameters were introduced by \citet{bender92}, who define a well-suited orthogonal coordinates in the three-space of the observable parameters.
Then, these authors compared the structural properties of ellipticals, bulges, compact ellipticals and dwarf ellipticals (dE) as described in the
$\kappa$-space, and found that all types of hot stellar systems, except for the very low luminosity dwarf spheroidals, defined planes in this
space with possible small, parallel offsets between classes. \citet{mclau08} defined the `$\kappa$' parameters using mass rather than
luminosity surface density, as this is more useful for comparing GCs to younger clusters and galaxies. In order to emphasize mass, \citet{mclau08}
used the notation $\kappa_{m,i}$:

\begin{equation}
\kappa_{m,1}=(\log \sigma_{p,0}^2 + \log R_{\rm h})/\sqrt{2}
\end{equation}
\begin{equation}
\kappa_{m,2}=(\log \sigma_{p,0}^2 + 2\log \Sigma_{\rm h} - \log R_{\rm h})/\sqrt{6}
\end{equation}
\begin{equation}
\kappa_{m,3}=(\log \sigma_{p,0}^2 - \log \Sigma_{\rm h} - \log R_{\rm h})/\sqrt{3}
\end{equation}

In calculating $\kappa_{m,1}$, $\kappa_{m,2}$, and $\kappa_{m,3}$, the $\sigma_{p,0}$ predicted in Table 10 (Column 11) by adoption of
population-synthesis mass-to-light ratios has been used, and $\Sigma_{\rm h}$ evaluated by adding $\log\,\Upsilon_V^{\rm pop}$ to $\log\,I_{\rm h}$
from Column (13) of Table 9. This is evident that $\kappa_{m,3}$ is independent of the assumed mass-to-light ratio. $R_{\rm h}$ is taken from
Table 9 (Column 7) but is in units of kpc rather than pc, for compatibility with the galaxy-oriented definitions of Bender et al. (1992).

\begin{table*}
\begin{minipage}{170mm}
\scriptsize
\caption{$\kappa-$space parameters of GC-1 and GC-2 in the M81 group.}\label{table11}
\begin{tabular}{ccccrcr}
\hline
ID & Detector & Band & Model & $\kappa_{m,1}$ & $\kappa_{m,2}$  & $\kappa_{m,3}$ \\
 &    &   &   &  &   & \\
(1) &  (2)  &  (3) &  (4) & (5) &  (6)  & (7) \\
\hline
 GC-1       & ACS/WFC    & 606        & K66      & $-0.487  ^{+0.149   }_{-0.161  }$ & $4.535   ^{+0.368   }_{-0.249  }$ & $0.112   ^{+0.095   }_{-0.110  }$\\
          &            &            & W        & $-0.305  ^{+0.122   }_{-0.142  }$ & $4.395   ^{+0.263   }_{-0.297  }$ & $0.197   ^{+0.072   }_{-0.083  }$\\
          &            &            & S        & $-1.442  ^{+0.166   }_{-0.183  }$ & $4.096   ^{+0.266   }_{-0.216  }$ & $-0.677  ^{+0.086   }_{-0.104  }$\\
        & ACS/WFC    & 814        & K66      & $-0.525  ^{+0.139   }_{-0.156  }$ & $4.524   ^{+0.355   }_{-0.243  }$ & $0.080   ^{+0.086   }_{-0.106  }$\\
          &            &            & W        & $-0.301  ^{+0.184   }_{-0.160  }$ & $4.384   ^{+0.449   }_{-0.354  }$ & $0.200   ^{+0.124   }_{-0.098  }$\\
          &            &            & S        & $-1.426  ^{+0.177   }_{-0.188  }$ & $4.120   ^{+0.243   }_{-0.196  }$ & $-0.663  ^{+0.095   }_{-0.108  }$\\
 GC-2       & WFC3/UVIS  & 606        & K66      & $0.055   ^{+0.086   }_{-0.080  }$ & $4.654   ^{+0.167   }_{-0.157  }$ & $0.247   ^{+0.048   }_{-0.038  }$\\
          &            &            & W        & $0.193   ^{+0.091   }_{-0.122  }$ & $4.476   ^{+0.179   }_{-0.234  }$ & $0.310   ^{+0.048   }_{-0.069  }$\\
          &            &            & S        & $-0.797  ^{+0.125   }_{-0.136  }$ & $4.214   ^{+0.189   }_{-0.177  }$ & $-0.440  ^{+0.055   }_{-0.065  }$\\
        & WFC3/UVIS  & 814        & K66      & $0.061   ^{+0.100   }_{-0.081  }$ & $4.754   ^{+0.214   }_{-0.140  }$ & $0.252   ^{+0.060   }_{-0.040  }$\\
          &            &            & W        & $0.164   ^{+0.083   }_{-0.085  }$ & $4.676   ^{+0.153   }_{-0.122  }$ & $0.286   ^{+0.041   }_{-0.036  }$\\
          &            &            & S        & $-0.798  ^{+0.127   }_{-0.134  }$ & $4.217   ^{+0.190   }_{-0.180  }$ & $-0.441  ^{+0.056   }_{-0.064  }$\\
\hline
\end{tabular}
\end{minipage}
\end{table*}


\begin{figure}
\resizebox{\hsize}{!}{\rotatebox{-90}{\includegraphics{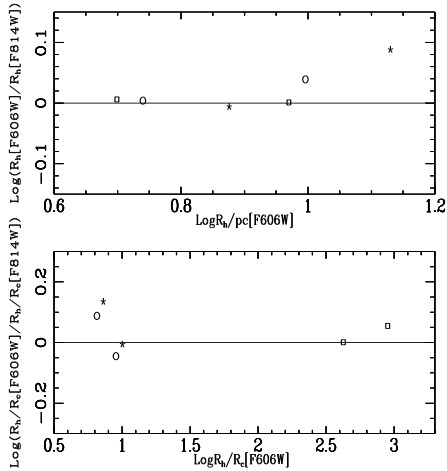}}}
\caption{Comparison of structural parameters for model fits to the sample clusters observed in both $F606W$ and $F814W$ bands. From top
to bottom: projected half-light radius and ratio of half-light to core radius. Open circles: \citet{king66} model; squares:
\citet{sersic68} model; stars: \citet{wilson75} model.}
\label{fig7}
\end{figure}

\begin{figure}
\centerline{\hfil
   \includegraphics[width=82mm]{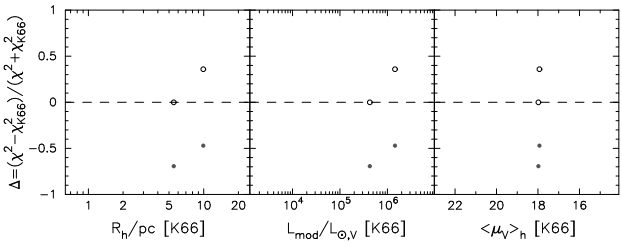}
\hfil}
\caption{Relative quality of fit for Wilson and S\'{e}rsic models (filled and open circles) versus King model for the sample clusters in
this paper.
\label{fig:fig8}
}
\end{figure}

\subsection{Comparison of fits in the $F606W$ and $F814W$ bands}

In this paper, we derived the structural and dynamical parameters from fitting three different models (King 1966, Wilson 1975
and S\'{e}rsic 1968) to the radial surface brightness profiles of GC-1 and GC-2 in the $F606W$ and $F814W$ bands. Assessment of the
systematic errors and colour dependencies in the fits can be done by comparing model fits to the same cluster observed in different
bands \citep{barmby07}. Fig. 7 compares the parameters derived from fits to GC-1 and GC-2 observed in both the $F606W$ and $F814W$ bands.
It is obvious that the agreement is quite good, with somewhat larger scatter for the \citet{wilson75} model fits to GC-2 in the
projected half-light radius. For both the projected half-light radius and ratio of half-light to core radius, the agreement is
excellent for the \citet{sersic68} model fits to both GC-1 and GC-2. \citet{barmby07} presented that the agreement is better
for the \citet{king66} model fits than the \citet{wilson75} and \citet{sersic68} model fits to the M31 star clusters. The results of
\citet{ma15} for the M33 star clusters is in agreement with those presented by \citet{barmby07} for the M31 star clusters. However,
we want to mention that there are only two sample star clusters here. The conclusion obtained here may not be correct for the M81
group star clusters.

\subsection{Comparison of three model fittings}

In order to clearly present the quality of fit for different models, we shows the relative quality of fit, $\Delta$, for the
Wilson- and S\'{e}rsic-model fits (filled and open circles, respectively, in all panels) versus the King-model fits for GC-1 and
GC-2 observed in the $F606W$ band in Fig. 8. $\Delta$ is defined as
\begin{equation}
\Delta=(\chi^2_{\rm alternate}-\chi^2_{\rm K66})/
(\chi^2_{\rm alternate}+\chi^2_{\rm K66})
\end{equation}
for comparing the $\chi^2$ of the best fit of an `alternate' model with the $\chi^2$ of the best fit of the King model
\citep[see][for details]{barmby07,mclau08}. If the parameter $\Delta$ is zero, the two models fit the same cluster equally well.
Positive values indicate a better fit of the King model, and negative values indicate the `alternate' model is a better fit than
the King model.

The $\Delta$ values are plotted as a function of some structural parameters, including the half-light radius $R_{\rm h}$, total model
luminosity $L_{\rm mod}$, and the intrinsic average surface brightness $\langle\mu_V\rangle_{\rm h}$. Fig. 8 shows that
GC-1 and GC-2 are significantly better fitted by the \citet{wilson75} model than by the \citet{king66} or \citet{sersic68} models
(GC-1 has $\Delta < -0.5$ for the Wilson-model fits and GC-2 has $\Delta \sim -0.5$ for the Wilson-model fits.). In addition, the
\citet{king66} model fits the data of GC-1 and GC-2 somewhat better than the \citet{sersic68} model. These results differ from those
for the clusters in NGC 5128 \citep{mclau08}, and in M31 \citep{barmby07,wang13} and M33 \citep{ma15}. \citet{mclau08} showed that
for the clusters in NGC 5128, the \citet{wilson75} model fits as well as or better than the \citet{king66} model; however, \citet{barmby07}
and \citet{wang13} showed that for most clusters in M31, the \citet{king66} model fits better than the \citet{wilson75} model. \citet{ma15}
indicated that the \citet{king66} and \citet{wilson75} models fit equally well for nearly all of the M33 sample clusters, and that the
\citet{wilson75} model fits the M33 sample cluster data better than the \citet{sersic68} model. \citet{roman12} indicated that the M33
young clusters (${\rm log~age < 8}$) are notably better fitted by the Elson-Freeman-Fall model \citep{elson87} with no radial truncation,
while the M33 older clusters show no significant differences between the \citet{king62} and the Elson et al. (1987) fits. In addition, we want
to mention that \citet{mm05} showed that globulars and young massive clusters in the Milky Way, Magellanic Clouds and Fornax dSph are
systematically better fitted by the \citet{wilson75} model than by the \citet{king66} model, and that the \citet{sersic68} model often
fits these globulars and young massive clusters about as well as the \citet{wilson75} model but can be significantly worse. GC-1 and GC-2
are most luminous clusters in the local universe. For luminous clusters in M31 and NGC 5128, the results are different. For M31 clusters,
\citet{barmby07} showed that the bright clusters ($L_{\rm mod}>10^5{\rm L}_{\odot}$) are better fitted by the \citet{king66} model; however, for
NGC 5128 clusters, \citet{mclau08} showed that the bright clusters are generally fitted much better by the \citet{wilson75} and
\citet{sersic68} models. The results of GC-1 and GC-2 are somewhat different from those of M31 and NGC 5128 clusters. In fact,
\citet{barmby07} provided the implication that the \citet{king66} model is more strongly preferred for more luminous M31 clusters.
They suggested that the preference for the \citet{king66} over the \citet{wilson75} model is due to some more subtle feature of the
observational data that have not yet been isolated. However, the conclusion of \citet{mclau08} is that the haloes of GCs in NGC 5128
are generically more extended than the \citet{king66} model allows. For GC-1 and GC-2, we consider that they have more extended haloes
than the \citet{king66} model allows.

\section{Comparison to Previous Results and Discussion}

\citet{Jang12} estimated the ages and masses of GC-1 and GC-2 based on the SDSS $ugriz$ data and on the SSP models of BC03.
They reported that they derived the ages of GC-1 and GC-2 to be $\sim 15.8$ Gyr, and the masses to be $\sim 2.51\times 10^6~\rm M_\odot$ for GC-1, and
$\sim 7.08\times 10^6~\rm M_\odot$ for GC-2. In this paper, we accurately redetermined the ages and masses of GC-1 and GC-2, based on the improved
data and sophisticated fitting methods. In particular, we used the photometric measurements of the 2MASS, which can partially break
the age-metallicity degeneracy, in combination with the NVU and optical photometry; and we used the magnitudes in the $UBVRI$ and
$JHK_{\rm s}$ bands for reducing mass uncertainties resulting from photometric uncertainties based on only the magnitudes in one
band. The ages and masses of GC-1 and GC-2 obtained by \citet{Jang12} and in this paper show that they are old and massive star clusters
in the local universe. In particular, their ages show that they are as old as the universe.

\citet{Jang12} presented the radial surface brightness profiles of GC-1 and GC-2 in the $F814W$ band, and they derived the core and
half-light radii of GC-1 and GC-2 by fitting the photometric data by the King model and the model of Elson et al. (1987) without
considering the PSF. As we know, the apparent core structures of GCs are strongly influenced by the PSF (see Barmby et al. 2007
and Mclaughlin et al. 2008 for details). So, it is needed that we redetermined the core radii of GC-1
and GC-2 considering the PSF. In addition, in order to study the properties of GC-1 and GC-2 for details in the
future, the other structural and dynamical parameters should be obtained besides the core and half-light radii. So, it is also needed
that we derived a wide range of structural and dynamical parameters of GC-1 and GC-2 considering the PSF (see Section 5.2
for details). The structural parameters include the central surface brightness and central potential, concentration indices, core
and half-light (effective) radii and total luminosity, and the predicted dynamical parameters include the internal velocity
dispersion, total mass, binding energy, central mass density, central escape velocity, relaxation time-scales and phase-space
densities. From the King model fits, \citet{Jang12} derived the core radii of GC-1 and GC-2 to be $0.0755\pm0.0003$ and
$0.1269\pm0.0003$ arcsec, respectively. By fitting the King model \citep{king66}, the core radii of GC-1 and GC-2 obtained here
to be $0.0316\pm0.002$ and $0.0891\pm0.005$ arcsec, respectively, based on the radial surface brightness profiles in the $F814W$
band. It is seen that the core radii obtained here are smaller than those of \citet{Jang12}. We argue that the difference
between the core radii obtained here and those obtained by \citet{Jang12} resulted from the PSF, i.e. \citet{Jang12} derived
this parameter without considering the PSF. In addition, by fitting the King model \citep{king66}, the half-light radii obtained
here are $5.45\pm0.22$~pc and $9.06\pm0.21$~pc, respectively, based on the radial surface brightness profiles in the $F814W$ band,
which are in good agreement with those of \citet{Jang12}, who derived the half-light radii of GC-1 and GC-2 to be $6.13\pm0.01$~pc
and $9.81\pm0.01$~pc, respectively. However, we should mention that the results of this paper showed that GC-1 and GC-2 are significantly
better fitted by \citet{wilson75} model than by \citet{king66} or \citet{sersic68} models. So, we use the half-radii obtained based
on the fitting by \citet{wilson75} model in Fig. 9.

Some authors derived the masses of the most massive Local Group clusters, such as 037--B327 [${M} \sim 8.5 \times 10^6~\rm M_\odot$
\citep{bk02} or ${M} \sim 3.0 \pm 0.5 \times 10^7~\rm M_\odot$ \citep{ma06a}] and G1 [${M} \sim (7-17)\times 10^6~\rm M_\odot$
\citep{meylan01} or ${M} \sim (5.8-10.6)\times10^6~\rm M_\odot$ \citep{ma09b}] in M31 and
$\omega$ Cen [${M} \sim (2.9 - 5.1) \times 10^6~\rm M_\odot$ \citep{meylan02}] in the Milky Way. \citet{MH04} derived the
masses of the most luminous GCs in NGC 5128 to range from $M \sim 1.0$ to $M \sim 9.0\times 10^6~\rm M_\odot$. Recently,
\citet{Mayya13} derived the mass of the brightest GC in M81, GC1, to be $M\sim 1.0\times10^7~\rm M_\odot$. Comparing with
these most massive GCs mentioned above, GC-1 and GC-2 are among the most massive clusters in the local universe.
In particular, GC-2 is one of the most massive clusters in the local universe. \citet{meylan01} argued that the very
massive GCs blur the former clear (or simplistic) difference between GCs and dwarf galaxies. In fact, \citet{Zinn98}
had proposed that the nucleated dEs would contribute their naked nuclei as a population of GCs when they
were accreted and disrupted by a larger galaxy. \citet{MH04} posited that some of the massive GCs in NGC 5128 are nuclei
of tidally dwarfs based on their large masses and the possible detection of `extratidal light' by \citet{harris02}
(but see McLaughlin et al. 2008, for counter-arguments for the massive GCs in NGC 5128). Based on the position of a cluster
on the size-luminosity plane, \citet{mackey05} suggested that the most luminous Local Group clusters, such as M54 and
$\omega$ Centauri in the Milky Way, G1 in M31 and the most luminous clusters in NGC 5128, are the cores of former dwarf
galaxies. The large masses of GC-1 and GC-2 may make them different from the rest of the GC population in M81 and M82.
\citet{Jang12} suggested that these two massive clusters are remnants of dwarf galaxies based on the empirically established
relation ${\rm log}r_{\rm h}=0.25\times M_V+2.95$ \citep{mackey05}, which forms the upper envelope of Galactic GCs in the $M_V$
versus $r_{\rm h}$ plane. In Section 5.3, we see that, for both GC-1 and GC-2, the fitted King model \citep{king66} falls clearly
below the measured profiles at the largest radii, although the model fit was accurate at smaller radii. In fact,
\citet{harris02} have already found this case for the most luminous GCs in NGC 5128, in which they explained that these
clusters exhibit `extratidal light' continuing outward past the nominal tidal radius obtained based on the \citet{king66}
model. These authors further suggested the `extratidal light' as the residual trace of the field-star population of dwarf
satellite galaxies, and these most luminous GCs as the luminous compact nuclei of former dwarf satellite galaxies. The large
masses and the possible detection of `extratidal light' for GC-1 and GC-2 may present the further evidence that these two
GCs are nuclei of tidally dwarfs as suggested by \citet{Jang12}. However, we want to mention the fact that, from Fig. 6,
we can see that the \citet{wilson75} model can fit the entire profiles of GC-1 and GC-2 accurately without the need to invoke
amounts of `extratidal light' implied by the \citet{king66} model fits. In fact, \citet{mclau08} argued that the `excess'
light implied by the \citet{king66} model fits to some massive clusters of NGC 5128 is not likely the signature of genuine tidal
debris, but a symptom of generic shortcomings in the model itself, i.e. the theoretical basis for it is weak in the low-density
and unrelaxed farthest reaches of cluster envelopes, and theses authors showed that the \citet{wilson75} model can fit these
massive clusters of NGC 5128 accurately. The results and comments of \citet{mclau08} may imply that these massive clusters in
NGC 5128 have not the residual trace of the field-star population of dwarf satellite galaxies. So, it is needed to be confirmed
in the future that the massive clusters of NGC 5128 are the luminous compact nuclei of former dwarf satellite galaxies. We argued
that GC-1 and GC-2 are in this case. In addition, GC-1 and GC-2 are not the most luminous clusters in M81 and M82 (see Fig. 9).
Especially, nearly 20 clusters are more luminous than GC-1. So, we suggested that GC-1 is only a luminous cluster in M81
and M82; however, GC-2 may be different from conventional clusters in M81 and M82. We will discuss GC-2 in detail below. From the
\citet{wilson75} model fits, the half-light radii of GC-2 obtained here are 13.49 and 11.02 pc based on the radial surface
brightness profiles in the $F606W$ and $F814W$ bands, respectively. Its large size and high luminosity ($M_V=-10.63$ mag) makes GC-2
to be an ultra-compact dwarfs (UCD) in the M81 group. Fig. 9 shows the half-light radius ($R_{\rm h}$) versus absolute magnitude ($M_V$)
diagram, where data for GC-1 and GC-2 are plotted along with those for the GCs (good and spectroscopically confirmed) in M81
\citep{N11} and the star clusters (older than 1 Gyr) in M82 (Lim et al. 2013), and for the UCDs from \citet{Brodie11}.
In Fig. 9, we used the half-light radii of GC-1 and GC-2 obtained by fitting the radial surface brightness profiles in the $F606W$ band
by the \citet{wilson75} model. Fig. 9 makes it evident that GC-2 occupies the same area of the UCDs, indicating that GC-2 is a UCD in the
M81 group. It is well known that UCDs are compact stellar systems that are more luminous and larger than typical GCs, but more compact
than typical dwarf galaxies. They were originally discovered independently by \citet{Hilker99} and \citet{Drinkwater00} in spectroscopic
surveys of the centre of the Fornax cluster, and they have sizes of $10<r_{\rm h}<100$ pc and luminosities of $M_V<-9$ mag \citep{Liu15}.
Although UCDs have so far been found mostly in dense environments, at the centres of galaxy clusters, or near massive galaxies, they have
also been discovered in galaxy groups and even relatively isolated galaxies \citep[see][for a review]{zhang15}. \citet{Evstigneeva07}
discovered one definite and four possible UCD candidates in a photometric search in six galaxy groups. All of these UCDs are intergalactic
and not associated with any particular galaxy. Especially, the UCD in the Dorado Group \citep{Evstigneeva07} is an intragroup object like
GC-2 in the M81 group. As mentioned above, \citet{Jang12} considered GC-2 to be a remnant of a dwarf galaxy based on the distribution
in the $M_V$ versus $r_{\rm h}$ plane, which accreted to M81 and is receding now. Till now, there has been ongoing debate about the origin of UCDs.
A number of different origins of UCDs were brought up: (1) they are merely luminous, genuine GCs \citep{Mieske02,Murray09}; (2) they are the
products of the merger of super star clusters \citep{fk02,Bekki04}; (3) they are the stripped nuclei of dEs
\citep{Bekki01,Bekki03,Goerdt08}; (4) they are remnants of primordial compact galaxies \citep{Drinkwater04}. \citet{Brodie11} presented
that the M87 UCDs are predominantly stripped nuclei. At the same time, these authors consider it more probable that UCDs have their origins
as nuclei that have since been stripped by tidal forces. However, \citet{Mieske04}, \citet{Mieske06} and \citet{Mieske12} presented that the luminosity and size
distribution of UCDs shows a smooth transition to the regime of GCs. So, these authors suggested that, from a statistical point of view, there
is no need to invoke an additional formation channel based on the conclusion that the number counts of UCDs are fully consistent with them
being the bright tail of the GC population. In addition, \citet{Hasegan05} and \citet{Mieske08} presented that, the more massive UCDs show
evidence for higher dynamical mass-to-light ratios than those derived from stellar population modelling alone, suggesting that some massive
UCDs host a large amount of dark matter \citep[see also][]{Pandya16}. For one UCD (M60-UCD1) it has been shown that a massive black hole is
responsible for the elevated mass-to-light ratio \citep{Seth14}. Especially, \citet{Mieske13} presented two UCD formation channels: one is
a `globular cluster channel' which is important mainly for UCDs with $M \la 10^7~\rm M_\odot$ and another is tidal transformation of
massive progenitor galaxies which dominates for UCDs with $M \ga 10^7~\rm M_\odot$ and still contributes for lower UCD masses. It is
unfortunate that we cannot derive the spectroscopically observed velocity dispersion of GC-2 in this paper, i.e. we cannot derive the
dynamical mass of GC-2. In addition, it is true that the mass of GC-2 is smaller than $10^7 {\rm M}_\odot$. So, we cannot give a clear conclusion
of the origin of GC-2 here.

\begin{figure}
\resizebox{\hsize}{!}{\rotatebox{-90}{\includegraphics{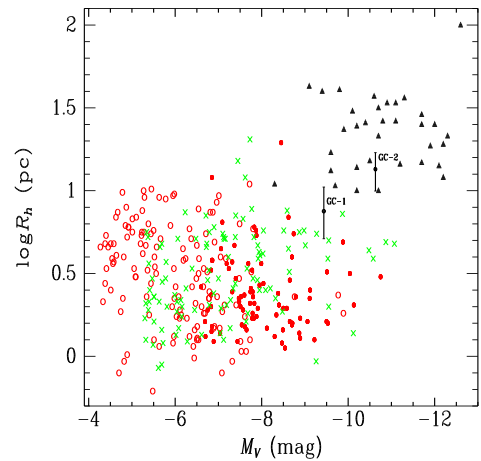}}}
\caption{Half-light radii vs. $M_V$ for GC-1 and GC-2 (black dots with error bars) in comparison with GCs (red dots) and GC candidates
(red circles) in M81 and star clusters (green crosses) in M82. The black filled triangles are the confirmed ultra-compact dwarfs (UCD) from \citet{Brodie11}.}
\label{fig9}
\end{figure}

\section{SUMMARY}

In this paper, we study the two GCs in the remote halo of M81 and M82 in the M81 group: GC-1 and GC-2.

(1) We derive the magnitudes in intermediate-band filters of the BATC system.

(2) We determine these two cluster's ages and masses by comparing their SEDs (from 2267 to 20000 {\AA}, comprising photometric data in the
near-ultraviolet of {\it GALEX}, 14 BATC intermediate-band and 2MASS near-infrared $JHK_{\rm s}$ filters) with theoretical stellar
population-synthesis models.

(3) The ages of GC-1 and GC-2 obtained in this paper are $15.50\pm3.20$ for GC-1 and $15.10\pm2.70$ Gyr, respectively, which showed that they
are old objects in the M81 group.

(4) The masses of GC-1 and GC-2 obtained in this paper are $1.77-2.04\times 10^6$ and $5.20-7.11\times 10^6~\rm M_\odot$, respectively,
which showed that they are among most massive clusters in the local universe, especially GC-2 is one of the most massive clusters in the
local universe.

(5) We derived the structural and dynamical parameters of GC-1 and GC-2 based on the {\it HST} images from fitting the surface brightness
profiles to three different models combined with the mass-to-light ratios ($M/L$ values) from population-synthesis models.

(6) Catalogues of fundamental structural and dynamical parameters obtained here are parallel in form to the catalogues produced by \citet{mm05}
for the GCs and massive young star clusters in the Milky Way, Magellanic Clouds, and the Fornax dwarf spheroidal, by \citet{barmby07} and
\citet{wang13} for the M31 star clusters, by \citet{ma15} for the M33 star clusters, and by \citet{mclau08} for the NGC 5128 star clusters.

(7) For the first time, we conclude that GC-2 is a UCD in the M81 group based on the $r_{\rm h}$ versus $M_V$ diagram.

\section*{Acknowledgments}
We would like to thank Dr. McLaughlin for his help in finishing this paper. He provide us a table including some parameters which being model-dependent function of $W_0$ or $c$.
We are indebted to the referee for his/her thoughtful comments and insightful suggestions that improved this paper greatly. This study has been supported by the National Natural
Science Foundation of China (NSFC, No. 11373035) and by the National Basic Research Program of China (973 Program, No. 2014CB845702). This study has also been supported by the
National Basic Research Program of China (973 Program, Nos. 2014CB845704 and 2013CB834902), by the National Natural Science Foundation of China (NSFC, Nos. 11603035,
11373033, 11433005, 11673027 and 11603034.



\end{document}